\documentclass[%
 reprint,
 aps,
 showpacs,
 amsmath,amssymb,
 dvipdfmx,
 longbibliograpy
]{revtex4-1}

\usepackage{graphicx}
\usepackage{dcolumn}
\usepackage{color}
\usepackage{bm}
\usepackage{here}
\usepackage{subfigure}
\usepackage{tabularx}
\newcolumntype{C}{>{\centering\arraybackslash}X}
\newcolumntype{L}{>{\raggedright\arraybackslash}X}
\newcolumntype{R}{>{\raggedleft\arraybackslash}X}
\usepackage{multirow}

\usepackage[
colorlinks=true, 
pdfstartview=FitV, 
linkcolor=blue, 
citecolor=magenta, 
urlcolor=blue, 
bookmarks=true,
bookmarksnumbered=true,
pdftitle={},
pdfauthor={}
]{hyperref}

\usepackage{amsthm}
\newtheorem{assumption}{Assumption}

\usepackage{comment}
\newcommand{\sectionprl}[1]{{\em #1}\/.---}

\begin{document}
\title{Long-range phase order in two dimensions under shear flow}
\author{Hiroyoshi Nakano$^1$, Yuki Minami$^2$, and Shin-ichi Sasa$^1$}
\affiliation{$^1$Department of Physics, Kyoto University, Kyoto 606-8502, Japan}
\affiliation{$^2$Department of Physics, Zhejiang University, Hangzhou 310027, China}

\date{\today}

\begin{abstract}
We theoretically and numerically investigate a two-dimensional O(2) model where an order parameter is convected by shear flow. We show that a long-range phase order emerges in two dimensions as a result of anomalous suppression of phase fluctuations by the shear flow. Furthermore, we use the finite-size scaling theory to demonstrate that a phase transition to the long-range ordered state from the disordered state is second order. At a transition point far from equilibrium, the critical exponents turn out to be close to the mean-field value for equilibrium systems.
\end{abstract}

\maketitle
\sectionprl{Introduction}
Nature exhibits various types of long-range order such as crystalline solids, liquid crystals, ferromagnets, and Bose--Einstein condensation. Whereas they are ubiquitous in the three-dimensional world, some types of long-range order associated with a continuous symmetry breaking are forbidden in two dimensions by the Mermin--Wagner theorem~\cite{Mermin1966,Hohenberg1967,Mermin1968}. The representative example of this theorem is that there is no long-range phase order in two dimensions.

Recently, the long-range phase order out of equilibrium has been attracted much attention. A stimulating example is the characteristic ``flocking" behavior among living things such as birds and bacteria. According to extensive numerical simulations of a simple model proposed by Vicsek \textit{et al.}~\cite{Vicsek1995}, the ``flocking" behavior was identified with the spontaneous emergence of the phase order in self-propelled polar particle systems - it is often called the polar order~\cite{Chate2020}. A remarkable feature here is that it occurs even in two dimensions~\cite{Toner1995,Toner1998,Nishiguchi2017,Dadhichi2018,Tanida2020,Prawar2020}, even though it is prohibited for equilibrium systems by the Mermin--Wagner theorem~\cite{Tasaki2020}. This phenomenon was also observed in the two-temperature conserved XY model~\cite{Bassler1995,Reichi2010}. It is now accepted that the long-range phase order can exist even in two dimensions for some non-equilibrium systems with short-range interactions.

The aim of this Letter is to clarify how the long-range phase order emerges in two dimensions under a small non-equilibrium perturbation to equilibrium systems. We study a two-dimensional O(2) model with short-range interaction. For equilibrium O(2) models, the dimension $d=2$ is marginal; specifically, the long-range phase order is broken by thermal fluctuations for $d\leq 2$, but is stable for $d>2$~\cite{Berezinskii1971,Kosterlitz1973,Kosterlitz1974,Koma1995}. Here, we impose infinitesimal shear flow on such a system and drive it into a non-equilibrium steady state. We then ask whether long-range phase order appears in the externally driven system. This Letter shows that the answer is yes and investigates its origin.

There is a long history of studying phase transitions driven by external non-equilibrium forces. Well-studied examples are related to the Ising universality class, such as critical fluids, binary mixtures and lattice gases~\cite{Katz1984,Beijeren1984,Wang1989,Leung1991,Caracciolo2004,Marro2005}. The phase transition under shear flow was one of the topic examined in this context~\cite{Onuki1979,Onuki1997,Onuki2002,Corberi1999,Cirillo2005,Hucht2009,Saracco2009,Winter2010,Sebastian2012}. As a seminal study, Onuki and Kawasaki performed the renormalization group analysis of the sheared critical fluids~\cite{Onuki1979}. Recently, some group studied the related systems by using Monte Carlo simulations~\cite{Cirillo2005,Saracco2009,Winter2010,Sebastian2012}.

Regarding externally driven systems with continuous symmetry, the main focus has been on three-dimensional phenomena such as an isotropic-to-lamellar transition of block copolymer melts~\cite{Cates1989,Koppi1993,Fredrickson1994,Zvelindovsky2000}, an isotropic-to-nematic transition of liquid crystals~\cite{Olmsted1990,Olmsted1992,Grizzuti2003,Lettinga2004,Hobbie2006,Ripoll2008}, an nematic-to-smectic transition of liquid crystals~\cite{DeGennes1976,Bruinsma1991,Safinya1991}, a crystallization of colloidal suspensions~\cite{Butler2002,Miyama2011}, and a spinodal decomposition of a large-$N$ limit model~\cite{Corberi2002,Corberi2003}. To our knowledge, the main question of this Letter has been never addressed before.

The key point of our study is to argue the stability of the long-range phase order in terms of the infrared divergence~\cite{Goldenfeld2018}. For the equilibrium O(2) model, the correlation function of the phase fluctuation behaves as $|\bm{k}|^{-2}$ where $\bm{k}$ represents the wavenumber. This fluctuation causes the logarithmic divergence of the real-space correlation function in the limit of large system size, and breaks the ordered state. Therefore, if stable long-range phase order appears under the shear flow, this logarithmic divergence must be removed by the flow effects. In this Letter, we theoretically demonstrate that the shear flow anomalously suppresses the phase fluctuation from $k_x^{-2}$ to $|k_x|^{-2/3}$, where the $x$-direction is defined as parallel to the flow. This new phase fluctuation is small enough to remove the divergence. Furthermore, by performing finite-size scaling analysis, we numerically show that the phase transition to the ordered state from the disordered state is second order. We also discuss our simulation result in the context of the previous results obtained for the sheared Ising model.

\sectionprl{Model}
Let $\bm{\varphi}(\bm{r},t) = (\varphi_1(\bm{r},t),\varphi_2(\bm{r},t))$ be a two-component real order parameter defined on a two-dimensional region $[0,L_x]\times[0,L_y]$. The order parameter is convected by the steady uniform shear flow with a velocity $\bm{v}(\bm{r}) = (\dot{\gamma} y,0)$, where $\dot{\gamma}\geq0$ without loss of generality. The dynamics is given by the time-dependent Ginzburg--Landau model:
\begin{eqnarray}
\Big[\frac{\partial}{\partial t} + \bm{v} \cdot \nabla  \Big] \varphi_a &=& - \Gamma \frac{\delta \Phi[\bm{\varphi}]}{\delta \varphi_a} + \eta_a,
\label{eq:model 1}\\
\langle \eta_a(t,\bm{r})\eta_b(t',\bm{r}') \rangle &=& 2 \Gamma T \delta_{ab} \delta(t-t')\delta(\bm{r}-\bm{r}') ,\label{eq:model 2}
\end{eqnarray}
where the Landau free energy $\Phi[\bm{\varphi}]$ is given by the standard $\varphi^4$ model
\begin{eqnarray}
\Phi[\bm{\varphi}] = \int d^2\bm{r} \Big[\frac{\kappa}{2} \sum_{a=1}^2(\nabla \varphi_a)^2 + \frac{r}{2} |\bm{\varphi}|^2 + \frac{u}{4}(|\bm{\varphi}|^2)^2\Big] .
\label{eq:model 3}
\end{eqnarray}
Here, $T$ is the temperature of the thermal bath and it is chosen independently of $r$.

The left-hand side of Eq.~(\ref{eq:model 1}) represents the rate of change following the flow. We stress that the convection term does not break the rotational symmetry in the order-parameter space. Furthermore, we note that in equilibrium our model is reduced to ``model A" in the classification of Hohenberg and Halperin~\cite{Hohenberg1977,Mazenko2008}. Because the steady-state distribution of $\bm{\varphi}$ is given by the canonical ensemble, the system exhibits quasi-long-range order instead of long-range order~\cite{Kosterlitz1973,Gupta1992}.

\sectionprl{Phase fluctuation in the low-temperature limit}
The state realized at $T=0$ is given by minimizing the Landau free energy $\Phi[\bm{\varphi}]$. For $r<0$, we have the ordered solution $\bm{\bar{\varphi}}=(\sqrt{-r/u},0)$, where we choose the direction of ordering as $\bm{n}=(1,0)$. In equilibrium, this ordered state is broken at finite temperature $T>0$. Here, we study how the shear flow suppresses the equilibrium fluctuations and stabilizes the ordered state in the low-temperature limit.

To analyze the fluctuations around $\bm{\bar{\varphi}}$, we transform the field variable as $\bm{\varphi}(\bm{r},t)=(\sqrt{-r/u}+A(\bm{r},t)) \big(\cos\theta(\bm{r},t), \sin\theta(\bm{r},t)\big)$, where $A(\bm{r},t)$ is the amplitude fluctuation and $\theta(\bm{r},t)$ the phase fluctuation. The phase fluctuation corresponds to the gapless mode associated with O(2) symmetry breaking~\cite{Goldstone1961,Nambu1961,Goldstone1962}. Therefore, we study the phase fluctuation below. Because the thermal fluctuations become sufficiently small in the low-temperature limit, we can neglect the periodicity of $\theta(\bm{r},t)$ and describe its dynamics within the linear approximation as
\begin{eqnarray}
\Big[\frac{\partial}{\partial t} - \dot{\gamma} k_x \frac{\partial}{\partial k_y} +\Gamma \kappa |\bm{k}|^2 \Big] \tilde{\theta} (\bm{k},t)= \tilde{\eta}_2(\bm{k},t),
\label{eq:NG mode: Fourier space}
\end{eqnarray}
where $\tilde{\theta}(\bm{k},t)$ is the Fourier transform of $\theta(\bm{r},t)$. The equal-time correlation function $C_{\theta \theta}(\bm{k})$ in the steady state is defined by $\big\langle \tilde{\theta} (\bm{k},t)\tilde{\theta} (\bm{k}',t)\big\rangle  = C_{\theta \theta}(\bm{k}) \delta(\bm{k}+\bm{k}')$, where $\langle \cdots \rangle$ represents the average in the steady state. From Eq.~(\ref{eq:NG mode: Fourier space}), $C_{\theta \theta}(\bm{k})$ is formally solved as~%
\footnote{
See Supplemental Material for detailed analysis of linear fluctuations.
}
\begin{eqnarray}
C_{\theta \theta}(\bm{k}) = T\Gamma \int_0^{\infty} ds e^{-\Gamma \kappa \big(s |\bm{k}|^2 + \frac{1}{2}\dot{\gamma} s^2k_xk_y + \frac{1}{12} \dot{\gamma}^2 s^3 k_x^2 \big)}
\label{eq: formal solution of NG mode}.
\end{eqnarray}
For $\dot{\gamma}=0$, Eq.~(\ref{eq: formal solution of NG mode}) is immediately integrated as $C_{\theta \theta}(\bm{k}) = T|\bm{k}|^{-2}/\kappa$. In two dimensions, the $|\bm{k}|^{-2}$ mode leads to the logarithmic divergence of the real-space correlation function and destroys the long-range order. 

For $\dot{\gamma}>0$, the asymptotic behavior of Eq.~(\ref{eq: formal solution of NG mode}) for small $\bm{k}$ is calculated as
\begin{eqnarray}
C_{\theta \theta}(\bm{k})  \simeq \frac{T}{c_0(\sqrt{\kappa} \dot{\gamma} |k_x|/\Gamma )^{{\frac{2}{3}}}+\kappa |\bm{k}|^2},
\label{eq: NG mode under shear flow}
\end{eqnarray}
where $c_0\simeq 2.04$. The shear flow stretches the fluctuations along the $x$-axis, which induces the anisotropic term $(\sqrt{\kappa} \dot{\gamma} |k_x|/\Gamma )^{{\frac{2}{3}}}$ in Eq.~(\ref{eq: NG mode under shear flow}). Because the exponent $2/3$ of this term is smaller than $2$, the equilibrium fluctuations are suppressed so that the logarithmic divergence in two dimensions is removed. Therefore, the fluctuations under shear flow do not break the long-range phase order for sufficiently low temperatures.

\begin{figure}[b]
\centering
\includegraphics[width=8.6cm]{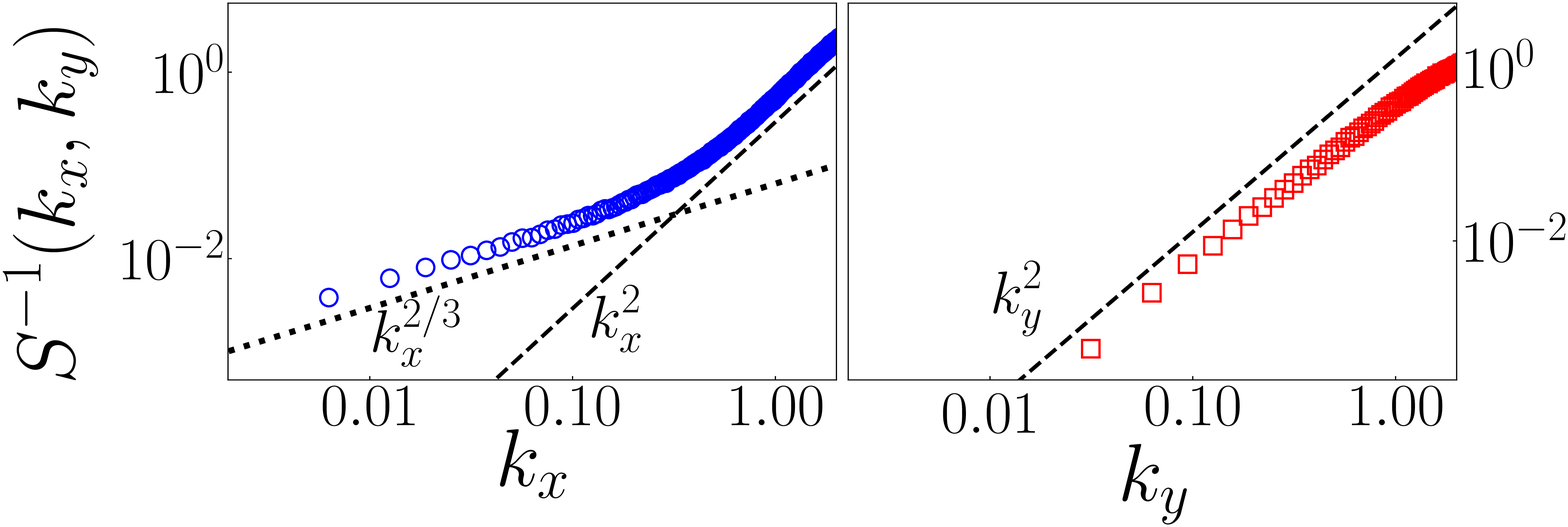} 
\caption{(Color online) Structure factor in ordered state. Left: $S^{-1}(k_x,k_y=0)$ versus $k_x$.  Right: $S^{-1}(k_x=0,k_y)$ versus $k_y$.}
\label{fig:plot of structure factor in ordered state}
\end{figure}

We also observe the exponent $2/3$ beyond the linear regime. To this end, we numerically solve the full equation (\ref{eq:model 1}) and calculate the structure factor, defined by $\langle \bm{\varphi}(\bm{k})  \cdot \bm{\varphi}(\bm{k}')\rangle = S(\bm{k}) \delta(\bm{k}+\bm{k}')$. Figure~\ref{fig:plot of structure factor in ordered state} plots $S^{-1}(\bm{k})$ for $\Gamma=T=u=1.0$, $\kappa=0.5$, $\dot{\gamma}=0.1$ and $r=-3.01$, where we have the long-range ordered state as explained below. From this figure, we find that the $k_x$-dependence of $S^{-1}(k_x,k_y=0)$ crosses over from $|k_x|^{-2/3}$ to $k_x^{-2}$. This behavior qualitatively agrees with the linearized model, Eq.~(\ref{eq: NG mode under shear flow}). 

We note that the length scale $l\equiv \sqrt{\kappa\Gamma /\dot{\gamma}}$ governs the crossover behavior. Because $l\to\infty$ in the equilibrium limit $\dot{\gamma} \to +0$, the order of the two limits $\bm{k} \to \bm{0}$ and $\dot{\gamma} \to +0$ cannot be exchanged. This observation leads to the result that the fractional mode $|k_x|^{-2/3}$ stabilizes the long-range order even when $\dot{\gamma} \to +0$.

\sectionprl{Finite-size scaling analysis}
We carry out finite-size scaling analysis to show further evidences of long-range order in the presence of shear flow.
Because the finite-size scaling theory in isotropic systems is modified by the anisotropy of shear flow~\cite{Winter2010}, we give an overview below of the finite-size scaling theory in the sheared system. Essentially the same analysis has been used for driven lattice gases~\cite{Binder1989,Wang1989,Leung1991,Caracciolo2004}.

The finite-size scaling theory is constructed on the basis of the scaling invariance at the second-order phase transition point. The scaling invariance is mathematically expressed by two relations. The first one is written using a free energy $F(\tau,h,L^{-1}_x,L^{-1}_y;\dot{\gamma})$ in the finite-size system, where $\tau=(r-r_c)/r_c$ is the dimensionless distance from the transition point $r_c$, and $h$ is the external field coupled with $\hat{m} = |\int d^2\bm{r} \bm{\varphi}(\bm{r})|/L_xL_y$. Then, the scaling invariance of the free energy near the critical point is given by the scaling relation
\begin{eqnarray}
\hspace{-0.5cm} F(\tau,h,L^{-1}_x,L^{-1}_y;\dot{\gamma}) = F(b^{z_\tau}\tau,b^{z_h}h,b^{z_{x}}L^{-1}_x,bL^{-1}_y;\dot{\gamma})
\label{eq:scaling assumption 1}
\end{eqnarray}
for any $b>0$, where the three scaling dimensions $z_{\tau}$, $z_h$, and $z_{x}$ are introduced. The second relation is that any quantity $\langle A \rangle_{h=0}$ in the absence of an external field can be expressed in terms of the correlation lengths $\xi_x$ and $\xi_y$ as
\begin{eqnarray}
\langle A \rangle_{h=0}(L^{-1}_x,L^{-1}_y,\tau;\dot{\gamma}) = L_x^{w_A}\mathcal{A}\Big(\frac{\xi_x}{L_x},\frac{\xi_y}{L_y};\dot{\gamma} \Big),
\label{eq:scaling assumption 2}
\end{eqnarray}
where $w_A$ is a constant and $\mathcal{A}$ is a scaling function. 

All the critical exponents are expressed by combinations of the three scaling dimensions $z_{\tau}$, $z_h$, and $z_{x}$~%
\footnote{See Supplemental Material for details of the finite-size scaling theory}.
For example, the critical exponents $\nu_x$ and $\nu_y$, characterizing the divergence of the correlation length (i.e. $\xi_i \sim |\tau|^{-\nu_i}$), are expressed as $\nu_x=z_{x}/z_{\tau}$ and $\nu_y=1/z_{\tau}$. The exponent $\beta$, characterizing the onset of magnetization slightly below the critical point (i.e. $\langle \hat{m}\rangle_{h=0} \sim |\tau|^{\beta}$), is given by $\beta=-\tilde{z}_h/z_{\tau}$, where we have introduced $\tilde{z}_h \equiv z_h-(1+z_x)$.

We note that $z_{x}$ characterizes the anisotropy of the divergence of the correlation length because it is rewritten as $\nu_x/\nu_y$.
Actually, the anisotropy of the shear flow makes $z_{x} \neq 1$. This can be immediately confirmed from the theoretical analysis of the linearized model by dropping the $\varphi^4$ term from Eq.~(\ref{eq:model 3}). This model is well-defined for $r>0$ and exhibits a singular divergence as $r \to +0$. From a similar calculation as the phase fluctuations in the low-temperature limit, we obtain $\nu_x=3/2$ and $\nu_y=1/2$, and then $z_{x}$ is given by $z_{x}=\nu_x/\nu_y=3$. Thus, it is natural to introduce $z_{x} \neq 1$ in the presence of the shear flow.

\begin{figure}[b]
\begin{center}
\includegraphics[width=8.6cm]{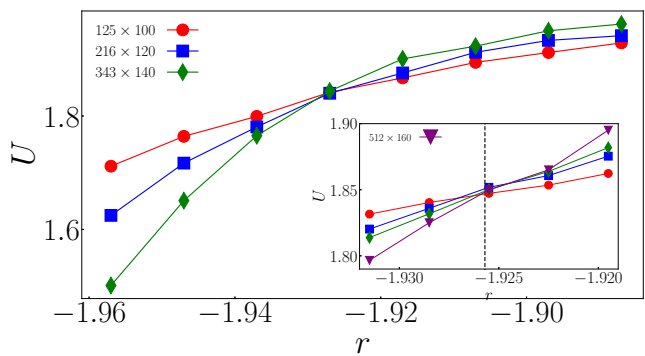}\\
\end{center}
\caption{(Color online) Binder parameter $U$ as a function of $r$ for $\dot{\gamma}=5.0$. Inset: zoom of the intersection point. The error bars of the data are in the order of the point sizes.}
\label{fig:plot of finite-size scaling for sr=5}
\end{figure}

Now, we show that the finite-size scaling theory works well for our model using numerical simulations. Below, we fix $\Gamma=T=u=1.0$ and $\kappa=0.5$, and treat $\dot{\gamma}$ and $r$ as control parameters. From Eqs.~(\ref{eq:scaling assumption 1}) and (\ref{eq:scaling assumption 2}), we can derive the system-size dependence of the $n$-th moment of magnetization as
\begin{eqnarray}
\hspace{-0.5cm} \big\langle \hat{m}^n \big\rangle_{h=0}(L^{-1}_x,L^{-1}_y,\tau;\dot{\gamma}) =  L_y^{\tilde{z}_h n} \mathcal{M}_n(L_y^{z_{\tau}}\tau,L_y^{z_{x}}L^{-1}_x;\dot{\gamma}).
\label{eq:system size dependence of kth cumulant}
\end{eqnarray}
The Binder parameter, defined by $U\equiv \big\langle \hat{m}^4 \big\rangle_{h=0}/\big\langle \hat{m}^2 \big\rangle_{h=0}^2$, satisfies
\begin{eqnarray}
U(L^{-1}_x,L^{-1}_y,{\tau};\dot{\gamma}) = \mathcal{U}(L_y^{z_{\tau}}\tau,L_y^{z_{x}}L^{-1}_x;\dot{\gamma}).
\label{eq:system size dependence of Binder parameter}
\end{eqnarray}
This equation means that all curves of the Binder parameter with different $L_x$ values intersect at a unique point when $L_y^{z_{x}}L^{-1}_x$ is fixed. In Fig.~\ref{fig:plot of finite-size scaling for sr=5}, we plot the numerical result for the Binder parameter $U$ for $\dot{\gamma}=5.0$. We have assumed $z_{x}=3$ with reference to the linearized model and chosen the system size as $L_x=125$, $216$, $343$, and $512$ under the condition $L_y=20L_x^{1/3}$. This figure shows the existence of the unique intersection point as expected.

\begin{figure*}[htb]
\begin{center}
\includegraphics[width=17.6cm]{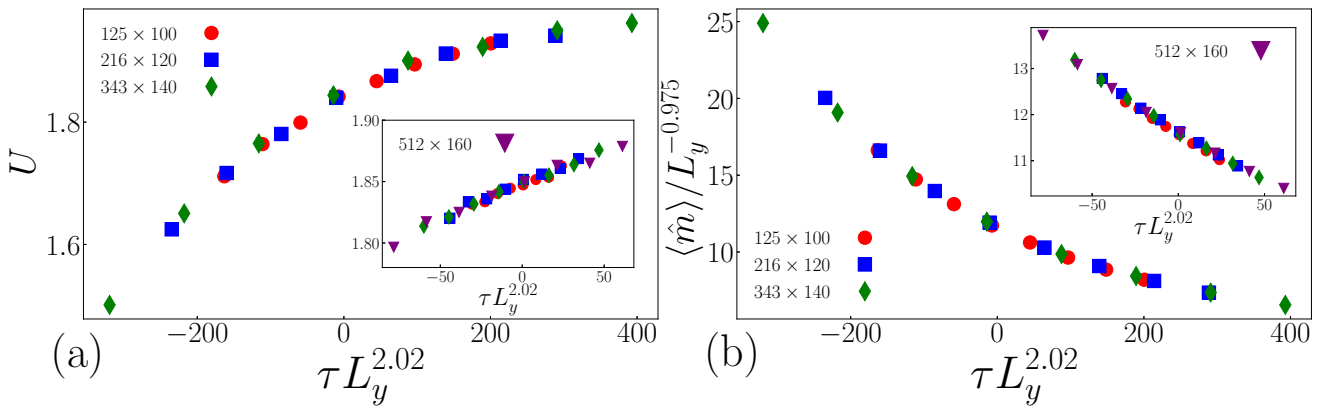}
\end{center}
\caption{(Color online) Finite-size scaling plot for $\dot{\gamma}=5.0$. (a): $U$ versus $\tau L_y^{z_{\tau}}$. (b) $\big\langle \hat{m} \big\rangle_{h=0} /L^{\tilde{z}_h}_y$ versus $\tau L_y^{z_{\tau}}$. In both figures, the inset is an enlargement of $\tau=0$. $r_c$, $z_{\tau}$ and $\tilde{z}_h$ are fixed at the best-fit value. The error bars of the data are in the order of the point sizes.}
\label{fig:rescale plot of finite-size scaling for sr=5}
\end{figure*}
According to the finite-size scaling relations Eqs.~(\ref{eq:system size dependence of kth cumulant}) and (\ref{eq:system size dependence of Binder parameter}), the magnetization $\big\langle \hat{m} \big\rangle_{h=0} 
$ and the Binder parameter $U$ can be expanded as power series near the critical point:
\begin{eqnarray}
\big\langle \hat{m} \big\rangle_{h=0} &=&  L_y^{\tilde{z}_h} \sum_{n=0}^{N} C^{m}_n(L_y^{z_{x}}L^{-1}_x) L_y^{z_{\tau} n}\tau^n,
\label{eq:system size dependence of kth cumulant: Taylor expansion}\\
U &=&  \sum_{n=0}^{N} C^{u}_n(L_y^{z_{x}}L^{-1}_x) L_y^{z_{\tau} n}\tau^n,
\label{eq:system size dependence of Binder parameter: Taylor expansion}
\end{eqnarray}
where $C^{m}_n$ and $C^{u}_n$ are expansion coefficients dependent on $L_y^{z_{x}}L^{-1}_x$. By fitting the simulation data to these expansions, we determine the critical point $r_c$ and the scaling exponent $(\tilde{z}_h,z_{\tau})$. In particular, we use the data in the region $-1.930<r<-1.920$ and perform simultaneous fitting of the two quantities $\big\langle \hat{m} \big\rangle_{h=0}$ and $U$ to Eqs.~(\ref{eq:system size dependence of kth cumulant: Taylor expansion}) and (\ref{eq:system size dependence of Binder parameter: Taylor expansion}) with $N=2$; we obtain $r_c=-1.9257\pm0.0002$, $z_{\tau}=2.05\pm0.11$, and $\tilde{z}_h=-0.983\pm0.026$. 
The validity of these fittings is shown in Fig.~\ref{fig:rescale plot of finite-size scaling for sr=5}, which is the scaled plot of the two quantities $\big\langle \hat{m} \big\rangle_{h=0}$ and $U$. The scaled data for the different system sizes overlap, verifying the finite-size scaling relations Eqs.~(\ref{eq:system size dependence of kth cumulant: Taylor expansion}) and (\ref{eq:system size dependence of Binder parameter: Taylor expansion}). It is noteworthy that the existence of the universal curve provides an evidence of $z_{x}=3$. We can also perform the consistency check of $z_{x}=3$ from the observation of $\nu_x$ and $\nu_y$ by using the property that $z_{x}$ is related to the anisotropy of the divergence of the correlation length~%
\footnote{
See Supplemental Material for supplemental numerical data.
}.

From the obtained values of $z_{\tau}$ and $\tilde{z}_h$, the critical exponent $\beta$ is calculated as $\beta=-\tilde{z}_h/z_{\tau} = 0.480 \pm 0.029$. This behavior is very similar to the result for the mean-field theory of the $\varphi^4$ model in equilibrium. It is consistent with the previous theoretical results for the sheared Ising model~\cite{Onuki1979,Hucht2009}, where the mean-field character is recovered under a sufficiently large shear rate or in the large limit.

\sectionprl{Phase diagram}
\begin{figure}[t]
\includegraphics[width=8.6cm]{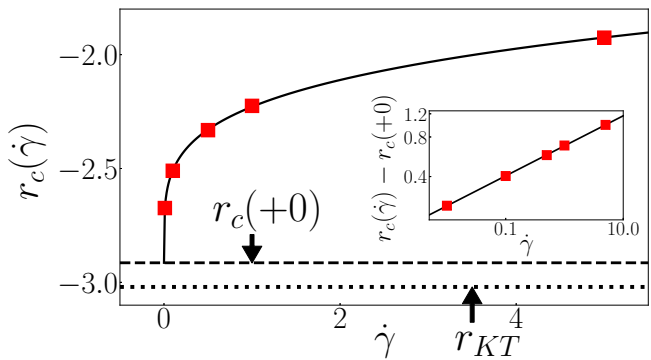} 
\caption{(Color online) Critical point $r_c$ as a function of $\dot{\gamma}$. The red points represent the numerical estimation and the black solid line Eq.~(\ref{eq:fitting function of critical point}) with the best-fit parameter.
Inset: $r_c-r_c(+0)$ versus $\dot{\gamma}$ with a log-log plot.}
\label{fig:phase diagram}
\end{figure}
We apply the above procedure to systems with smaller $\dot{\gamma}$ and show the phase diagram in Fig.~\ref{fig:phase diagram}, where the critical point $r_c$ is plotted as a function of $\dot{\gamma}$. For all $\dot{\gamma}$ values we have examined, the assumption $z_{x}=3$ is valid and the long-range phase order exists below $r_c$. We then ask where $r_c(\dot{\gamma})$ terminates as $\dot{\gamma} \to +0$. To answer this question, we assume that the critical point $r_c$ behaves as a function of $\dot{\gamma}$ in the form
\begin{eqnarray}
r_c(\dot{\gamma}) = D_0 \dot{\gamma}^{w} + r_c(+0).
\label{eq:fitting function of critical point}
\end{eqnarray}
Note that for the sheared Ising model, this functional form is known to reproduce the behavior of the critical point for small $\dot{\gamma}$~\cite{Onuki2002,Saracco2009,Winter2010,Sebastian2012}. By fitting the simulation data to Eq.~(\ref{eq:fitting function of critical point}), we obtain the best-fit parameters $r_c(+0)=-2.9139\pm 0.0151$, $D_0=0.685\pm0.016$, and $w = 0.228\pm0.006$. The corresponding curve is drawn as the black solid one in Fig.~\ref{fig:phase diagram}. The good agreement between the numerical estimation and the best-fit curve confirms the validity of Eq.~(\ref{eq:fitting function of critical point}) for our model. This gives the evidence that the long-range phase order is stabilized even under the infinitesimal shear flow. The critical point at the infinitesimal shear rate $\dot{\gamma}\to +0$ is estimated as $r_c(+0)=-2.9139\pm 0.0151$.

\sectionprl{Discussion}
We remember that our model exhibits the Kosterlitz--Thouless transition in equilibrium. The transition point is estimated to be $r_{KT}=-3.0204\pm0.0087$~\footnotemark[3]. Then, our results show that there exists a slight deviation between $r_c(+0)$ and $r_{KT}$. We here discuss two possible scenarios. The first one is that this deviation disappears by using the data at smaller $\dot{\gamma}$ for larger systems. Actually, for the two-dimensional Ising model~\cite{Saracco2009,Winter2010} and the three-dimensional critical fluid~\cite{Onuki2002}, $r_c(\dot{\gamma})$ terminates at the equilibrium transition point as $\dot{\gamma} \to +0$. As the second scenario, this deviation may remains for larger systems because the long-wavelength fluctuations ($|k_x|<2\pi/l$) are drastically altered even when $\dot{\gamma} \to +0$ (See Fig.~\ref{fig:plot of structure factor in ordered state}). Which scenario is correct is left for future study.

Another question for small $\dot{\gamma}$ is about the critical exponent $\beta$. Our simulation showed that for $\dot{\gamma}=0.5$, $\beta$ agrees well with the mean-field value as in the case of $\dot{\gamma}=5.0$. In contrast, for $\dot{\gamma}=0.01$, we obtained $z_{\tau}=2.03\pm 0.30$ and $\tilde{z}_h=-0.579\pm 0.02$, which corresponds to $\beta = 0.285$~\footnotemark[3]. Clearly, there is a large deviation between the observed result and mean-field theory. We do not judge whether this deviation comes from the finite-size effects or remains in the large system-size limit. On a related note, this problem also remains controversial for the sheared Ising model~\cite{Hucht2009,Saracco2009,Winter2010,Sebastian2012}. More careful analysis for smaller $\dot{\gamma}$ is necessary.

The phase mode induced by the shear flow, $S(k_x,k_y=0)\sim |k_x|^{-2/3}$, and the long-range order are two sides of the same coin. The interesting point is that the exponent $2/3$ is numerically observed for all $\dot{\gamma}$ values we have examined, although it is derived without considering the nonlinear interaction of fluctuations. This observation suggests that nonlinear effects are irrelevant for the structure factor in the ordered state. The theoretical verification of this conjecture is left as a future work.

Finally, we discuss possible experiments associated with our result. The model in this Letter gives an ideal description of some experimental systems using liquid undercooled metals~\cite{Reske1995,Albrecht1997} and magnetic fluids~\cite{Nijmeijer1995}. The liquid undercooled metal is known to exhibit the liquid ferromagnet phase due to short-ranged exchange interactions in three dimensions~\cite{Reske1995,Albrecht1997}. We expect that the two-dimensional liquid ferromagnet phase can be observed by designing a two-dimensional system~\cite{Nishiguchi2017}.

\bigskip
\sectionprl{Acknowledgements}
We thank M. Kobayashi for helpful comments on the numerical simulation, D. Nishiguchi for a critical reading of the manuscript, and M. Hongo for stimulating conversations. HN and SS are supported by KAKENHI (Nos. 17H01148, 19H05795, and 20K20425). YM is supported by the Zhejiang Provincial Natural Science Foundation Key Project (Grant No. LZ19A050001) and NSF of China (Grants No. 11975199 and 11674283).

%


\setcounter{equation}{0}
\setcounter{figure}{0}
\setcounter{table}{0}
\setcounter{page}{1}
\renewcommand{\thepage}{S\arabic{page}}  
\renewcommand{\thesection}{S\arabic{section}}   
\renewcommand{\thetable}{S\arabic{table}}   
\renewcommand{\thefigure}{S\arabic{figure}}
\renewcommand{\theequation}{S\arabic{equation}}
\renewcommand{\bibnumfmt}[1]{[S#1]}
\renewcommand{\citenumfont}[1]{S#1}

\begin{widetext}
\begin{center}
{\large \bf Supplemental Material for  \protect \\ 
  ``Long-range phase order in two dimensions under shear flow" }\\
\vspace*{0.3cm}
Hiroyoshi Nakano$^{1}$, Yuki Minami$^{2}$, and Shin-ichi Sasa$^{1}$
\\
\vspace*{0.1cm}
$^{1}${\small \it Department of Physics, Graduate School of Science, Kyoto University, Kyoto, Japan} \\
$^{2}${\small \it Department of Physics, Zhejiang University, Hangzhou 310027, China} 
\end{center}

\section{Linear analysis of fluctuations}
We present the details of the linear analysis of fluctuations for the model Eqs.~(1), (2), and (3). We first derive, in Sec.~\ref{sec:Formal solution of linearized model}, the exact integral expression of correlation functions such as Eq.~(5). Next, in Sec.~\ref{sec:fluctuations near criticality}, by using this expression, we perform the asymptotic analysis of fluctuations near the critical point. This result is applied to estimate the value of $z_{x}$ in the main text. Finally, in Sec.~\ref{subsec:Fluctuations in ordered state}, we analyze linear fluctuations around the ordered solution. We also give the detail of derivation of Eq.~(6).

\subsection{Formal solution of linearized model}
\label{sec:Formal solution of linearized model}
The linearized model is obtained by dropping the non-linear term from Eqs.~(1), (2), and (3) as
\begin{eqnarray}
\frac{\partial \varphi_a}{\partial t}  + \dot{\gamma} y \frac{\partial \varphi_a}{\partial x} &=& - \Gamma \Big(-\kappa \Delta + r \Big) \varphi_a+ \eta_a.
\label{eq:explicit form of linearized order parameter dynamics}
\end{eqnarray}
By the Fourier transform
\begin{eqnarray}
\tilde{\varphi}_a(\bm{k}) \equiv \int d^2 \bm{r} \varphi_a(\bm{r}) e^{i\bm{k}\cdot \bm{r}},
\end{eqnarray}
Eq.~(\ref{eq:explicit form of linearized order parameter dynamics}) is rewritten as
\begin{eqnarray}
\frac{\partial \tilde{\varphi}_a}{\partial t}  - \dot{\gamma} k_x \frac{\partial \tilde{\varphi}_a}{\partial k_y} &=& - \Gamma \Big(\kappa|\bm{k}|^2 + r \Big) \tilde{\varphi}_a+ \tilde{\eta}_a.
\label{eq:explicit form of linearized order parameter dynamics in Fourier space}
\end{eqnarray}
In this subsection, we calculate $C_{\varphi\varphi}(\bm{k},t)$ defined by $\langle \tilde{\varphi}_a(\bm{k},t)\tilde{\varphi}_b(\bm{k}',t)\rangle = C_{\varphi\varphi}(\bm{k},t) \delta_{ab}\delta(\bm{k}+\bm{k}')$. Our argument is essentially the same as that by Onuki and Kawasaki~\cite{smOnuki1979}.

We first derive the equation for $C_{\varphi\varphi}(\bm{k},t)$. Multiplying Eq.~(\ref{eq:explicit form of linearized order parameter dynamics in Fourier space}) by $\tilde{\varphi}_b(\bm{k}',t)$ yields
\begin{eqnarray}
\tilde{\varphi}_b(\bm{k}',t)\frac{\partial}{\partial t} \tilde{\varphi}_a(\bm{k},t) + \Big(- \dot{\gamma} k_x \frac{\partial}{\partial k_y} + \Gamma \big(\kappa|\bm{k}|^2 + r \big) \Big)\tilde{\varphi}_a(\bm{k},t) \tilde{\varphi}_b(\bm{k}',t)  = \tilde{\eta}_a(\bm{k},t) \tilde{\varphi}_b(\bm{k}',t).
\label{eq:EOM of ETCF: unfinished}
\end{eqnarray}
By taking the average over the noise, we have
\begin{eqnarray}
\Big\langle \tilde{\varphi}_b(\bm{k}',t)\frac{\partial}{\partial t} \tilde{\varphi}_a(\bm{k},t) \Big\rangle+ \Big(- \dot{\gamma} k_x \frac{\partial}{\partial k_y} + \Gamma \big(\kappa|\bm{k}|^2 + r \big) \Big)\Big\langle\tilde{\varphi}_a(\bm{k},t) \tilde{\varphi}_b(\bm{k}',t)\Big\rangle  = \Big\langle\tilde{\eta}_a(\bm{k},t) \tilde{\varphi}_b(\bm{k}',t)\Big\rangle.
\label{eq:EOM of ETCF: unfinished 1}
\end{eqnarray}
Because $\big\langle \tilde{\eta}(\bm{k},t) \tilde{\varphi}_a(\bm{k}',t) \big\rangle$ is defined by the Stratonovich convention, we have
\begin{eqnarray}
\big\langle \tilde{\eta}_a(\bm{k},t) \tilde{\varphi}_b(\bm{k}',t) \big\rangle =  T\Gamma \delta_{ab}\delta(\bm{k}+\bm{k}'),
\end{eqnarray}
where $\delta(\bm{k})$ is the delta function with argument $\bm{k}$, and Eq.~(\ref{eq:EOM of ETCF: unfinished 1}) is rewritten as
\begin{eqnarray}
\Big\langle \tilde{\varphi}_b(\bm{k}',t)\frac{\partial}{\partial t} \tilde{\varphi}_a(\bm{k},t) \Big\rangle+ \Big(- \dot{\gamma} k_x \frac{\partial}{\partial k_y} + \Gamma \big(\kappa|\bm{k}|^2 + r \big) \Big)\Big\langle\tilde{\varphi}_a(\bm{k},t) \tilde{\varphi}_b(\bm{k}',t)\Big\rangle  = T\Gamma \delta_{ab}\delta(\bm{k}+\bm{k}').
\label{eq:EOM of ETCF: unfinished 2}
\end{eqnarray}
Here, by definition, the time derivative of $C_{\varphi\varphi}(\bm{k},t)$ satisfies the following equation: 
\begin{eqnarray}
\Big\langle \tilde{\varphi}_a(\bm{k},t)\frac{\partial \tilde{\varphi}_b(\bm{k}',t)}{\partial t} \Big\rangle + \Big\langle \frac{\partial \tilde{\varphi}_a(\bm{k},t)}{\partial t}\tilde{\varphi}_b(\bm{k}',t) \Big\rangle = \frac{\partial C_{\varphi\varphi}(\bm{k},t)}{\partial t} \delta_{ab}\delta(\bm{k}+\bm{k}').
\label{eq:EOM of ETCF: unfinished 3}
\end{eqnarray}
By substituting Eq.~(\ref{eq:EOM of ETCF: unfinished 2}) into Eq.~(\ref{eq:EOM of ETCF: unfinished 3}), we obtain
\begin{eqnarray}
\frac{\partial C_{\varphi\varphi}(\bm{k},t)}{\partial t} = \Big( \dot{\gamma} k_x \frac{\partial}{\partial k_y} - 2\Gamma \big(\kappa|\bm{k}|^2 + r \big) \Big) C_{\varphi\varphi}(\bm{k},t) + 2 T \Gamma .
\label{eq:EOM of ETCF: full plus time derivative}
\end{eqnarray}
This equation describes the time evolution of $ C_{\varphi\varphi}(\bm{k},t)$. Because we are especially interested in the steady-state correlation, we set the time derivative of Eq.~(\ref{eq:EOM of ETCF: full plus time derivative}) to zero and study the equation 
\begin{eqnarray}
\Big(-\frac{1}{2} \dot{\gamma} k_x \frac{\partial}{\partial k_y} + \Gamma \big(\kappa|\bm{k}|^2 + r \big) \Big)C_{\varphi\varphi}(\bm{k}) =T \Gamma ,
\label{eq:EOM of ETCF: full}
\end{eqnarray}
where $C_{\varphi\varphi}(\bm{k})$ is the steady state correlation defined by $C_{\varphi\varphi}(\bm{k}) \equiv \lim_{t \to \infty} C_{\varphi\varphi}(\bm{k},t)$.

In equilibrium, Eq.~(\ref{eq:EOM of ETCF: full}) takes the simple form
\begin{eqnarray}
\Gamma \big(\kappa|\bm{k}|^2 + r \big) C_{\varphi\varphi}(\bm{k}) =T \Gamma ,
\end{eqnarray}
and we immediately find that $C_{\varphi\varphi}(\bm{k})$ is expressed as
\begin{eqnarray}
 C_{\varphi\varphi}(\bm{k}) = \frac{T}{\kappa|\bm{k}|^2 + r} .
\end{eqnarray}
For $\dot{\gamma} \neq 0$, although the first term of Eq.~(\ref{eq:EOM of ETCF: full}) is the differential operator, we can write $C_{\varphi\varphi}(\bm{k})$ as
\begin{eqnarray}
 C_{\varphi\varphi}(\bm{k}) = \frac{\Gamma T}{ - \frac{1}{2}\dot{\gamma} k_x \frac{\partial}{\partial k_y} + \Gamma \big(\kappa|\bm{k}|^2 + r \big)} .
 \label{eq:transient of formal solution 1}
\end{eqnarray}
The inverse operator of Eq.~(\ref{eq:transient of formal solution 1}) is expressed as
\begin{eqnarray}
 \frac{1}{ - \frac{1}{2}\dot{\gamma} k_x \frac{\partial}{\partial k_y} + \Gamma \big(\kappa|\bm{k}|^2 + r \big)} = \int_0^{\infty} ds e^{-s \big\{ - \frac{1}{2}\dot{\gamma} k_x \frac{\partial}{\partial k_y} + \Gamma (\kappa|\bm{k}|^2 + r ) \big\}}.
\label{eq:transient of formal solution 2}
\end{eqnarray}
Here, it is known that the exponential operator containing the first-order differential is decomposed as
\begin{eqnarray}
e^{\lambda U(x) + \lambda \frac{\partial}{\partial x}} = e^{\int_0^{\lambda} d\lambda' U(x+\lambda')} e^{\lambda \frac{\partial}{\partial x}}.
\label{eq:onuki formula}
\end{eqnarray}
We will give a proof of this formula later. By applying this formula to Eq.~(\ref{eq:transient of formal solution 2}) and substituting it into Eq.~(\ref{eq:transient of formal solution 1}), we obtain
\begin{eqnarray}
C_{\varphi\varphi}(\bm{k}) = T\Gamma \int_0^{\infty} ds e^{-\int_0^s d\lambda \Gamma (\kappa|\bm{\kappa}_{\lambda}|^2 + r )}
\label{eq:transient of formal solution 3}
\end{eqnarray}
with
\begin{eqnarray}
\bm{\kappa}_{\lambda} = (k_x,k_y+\frac{1}{2}\dot{\gamma} \lambda k_x).
\end{eqnarray}
The $\lambda$-integral in Eq.~(\ref{eq:transient of formal solution 3}) is straightforwardly calculated as
\begin{eqnarray}
C_{\varphi\varphi}(\bm{k})  = T\Gamma \int_0^{\infty} ds e^{-\Gamma\big\{s(\kappa|\bm{k}|^2+r) + \frac{1}{2}\kappa\dot{\gamma} s^2k_xk_y + \frac{1}{12} \kappa\dot{\gamma}^2 s^3 k_x^2 \big\}}.
\label{eq:correlation function: final expression}
\end{eqnarray}
This is the desired expression for the correlation function under shear flow. This type of expression was firstly derived by Onuki and Kawasaki~\cite{smOnuki1979} and widely used in the analyses of fluctuations in the presence of shear flow~\cite{smDeGennes1976,smFredrickson1986,smCates1989,smCorberi2002,smCorberi2003,smWada2003,smWada2004,smOtsuki2009}.

\subsubsection*{Derivation of Eq.~(\ref{eq:onuki formula})}
We here give a proof of the formula Eq.~(\ref{eq:onuki formula}). We define two functions by
\begin{eqnarray}
g(x,\lambda) \equiv e^{\lambda U(x)+\lambda \frac{\partial}{\partial x}} f(x),\\[3pt]
h(x,y) \equiv g(x,y-x)
\end{eqnarray}
for any function $f(x)$. Noting
\begin{eqnarray}
\frac{\partial g(x,\lambda)}{\partial \lambda} =  U(x) g(x,\lambda) + \frac{\partial g(x,\lambda)}{\partial x},
\end{eqnarray}
we calculate the $x$-differential of $h(x,y)$ as
\begin{eqnarray}
\frac{\partial h(x,y)}{\partial x} = - U(x) h(x,y).
\end{eqnarray}
Then, by integrating with respect to $x$, we obtain
\begin{eqnarray}
h(x,y) &=& h(y,y) e^{-\int_{y}^{x}ds U(s)} \nonumber \\[3pt]
&=& h(y,y) e^{\int_{0}^{y-x}ds U(y-s)}.
\end{eqnarray}
This leads to
\begin{eqnarray}
g(x,\lambda) &=& h(x,x+\lambda) \nonumber \\[3pt]
&=& h(x+\lambda,x+\lambda)e^{\int_{0}^{\lambda}ds U(x+\lambda-s)}\nonumber \\[3pt]
&=& g(x+\lambda,0)e^{\int_{0}^{\lambda}d\lambda' U(x+\lambda')}.
\label{eq:proof of onuki formula: 1-1}
\end{eqnarray}
Because $e^{\lambda \frac{\partial}{\partial x}}$ is the translational operator, we have
\begin{eqnarray}
g(x+\lambda,0) = f(x+\lambda)= e^{\lambda \frac{\partial}{\partial x}} f(x) .
\label{eq:proof of onuki formula: 1-2}
\end{eqnarray}
By combining Eqs.~(\ref{eq:proof of onuki formula: 1-1}) and (\ref{eq:proof of onuki formula: 1-2}), we obtain the desired identity
\begin{eqnarray}
e^{\lambda U(x)+\lambda \frac{\partial}{\partial x}} f(x) = e^{\int_{0}^{\lambda}d\lambda' U(x+\lambda')}e^{\lambda \frac{\partial}{\partial x}} f(x).
\end{eqnarray}

\subsection{Fluctuations near the critical point}
\label{sec:fluctuations near criticality}
The linearized model (\ref{eq:explicit form of linearized order parameter dynamics}) or (\ref{eq:explicit form of linearized order parameter dynamics in Fourier space}) is valid for $r>0$, and exhibits a singular behavior as $r \to +0$. This singular behavior is characterized by the divergence of the correlation length. As the simplest example, let us consider the case $k_x=0$. In this case, we immediately calculate the $s$-integral in (\ref{eq:correlation function: final expression}) for any $r$ and obtain
\begin{eqnarray}
C_{\varphi\varphi}(k_x=0,k_y) = \frac{T}{r+\kappa k_y^2}.
\end{eqnarray}
Then, we find that the correlation length $\xi_y$ diverges as
\begin{eqnarray}
\xi_y \sim r^{-1/2}
\end{eqnarray}
as $r\to +0$. Except for the simple wavelength region, we cannot perform the $s$-integral in (\ref{eq:correlation function: final expression}). Then, we focus on the asymptotic behavior of $C_{\varphi\varphi}(\bm{k})$ in the long-wavelength region and argue how the correlation length diverges. Below, $r$ is assumed to be sufficiently small.

We study two limiting $\bm{k}$-regions:
\begin{eqnarray}
{\rm (i)} \ \ \frac{1}{12} \dot{\gamma}^2 k_x^2 \ll \Gamma^2 \big(\kappa |\bm{k}|^2+r\big)^3, \\
{\rm (ii)} \ \  \frac{1}{12} \dot{\gamma}^2 k_x^2 \gg \Gamma^2 \big(\kappa |\bm{k}|^2+r\big)^3.
\end{eqnarray}
The schematic image of each region is drawn in Fig.~\ref{fig:dividing into two regions}. Naively speaking, region (i) is located near $k_x=0$, and region (ii) corresponds to all other region.
\begin{figure}[b]
\centering
\includegraphics[width=6cm]{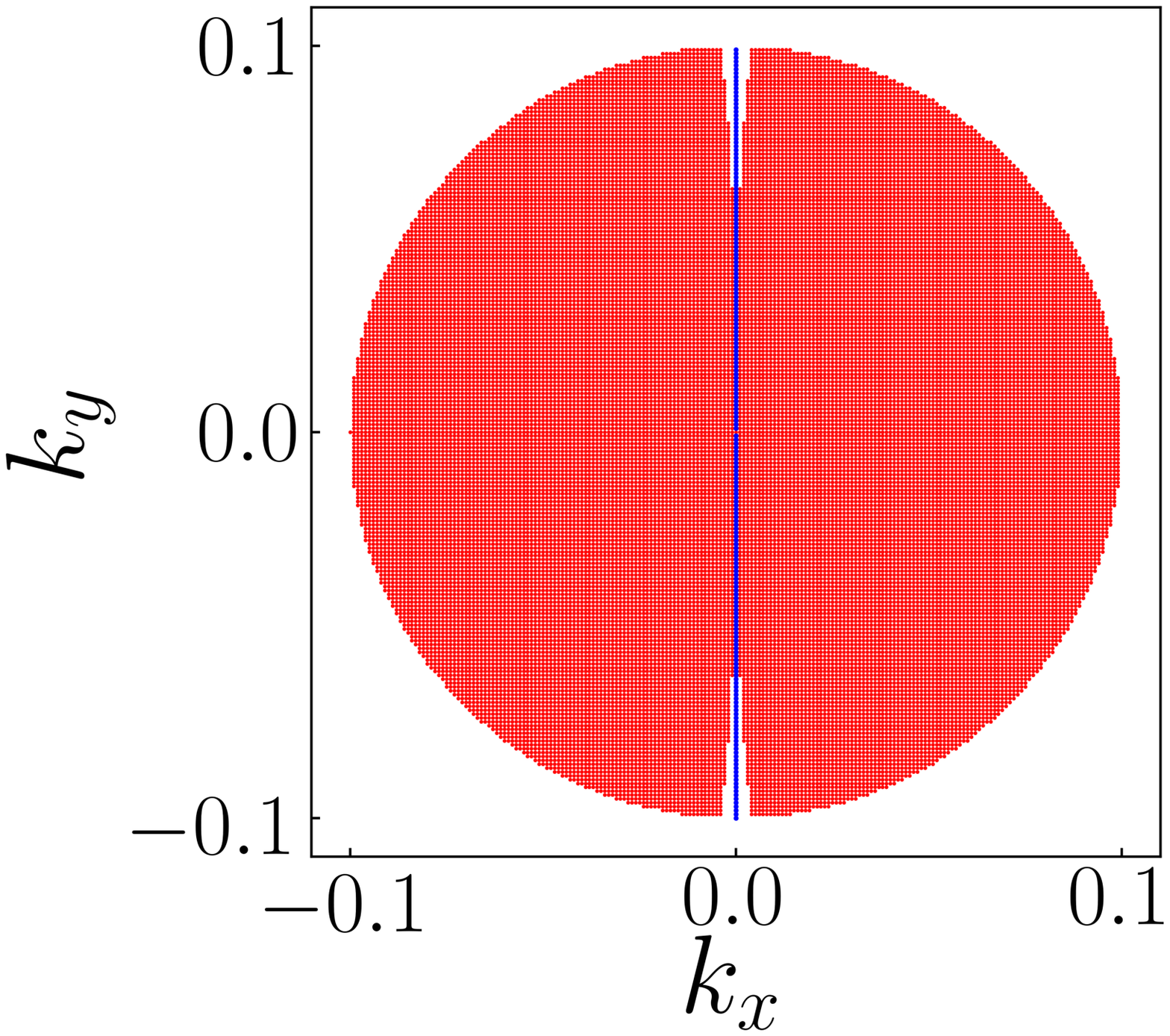} 
\includegraphics[width=5.5cm]{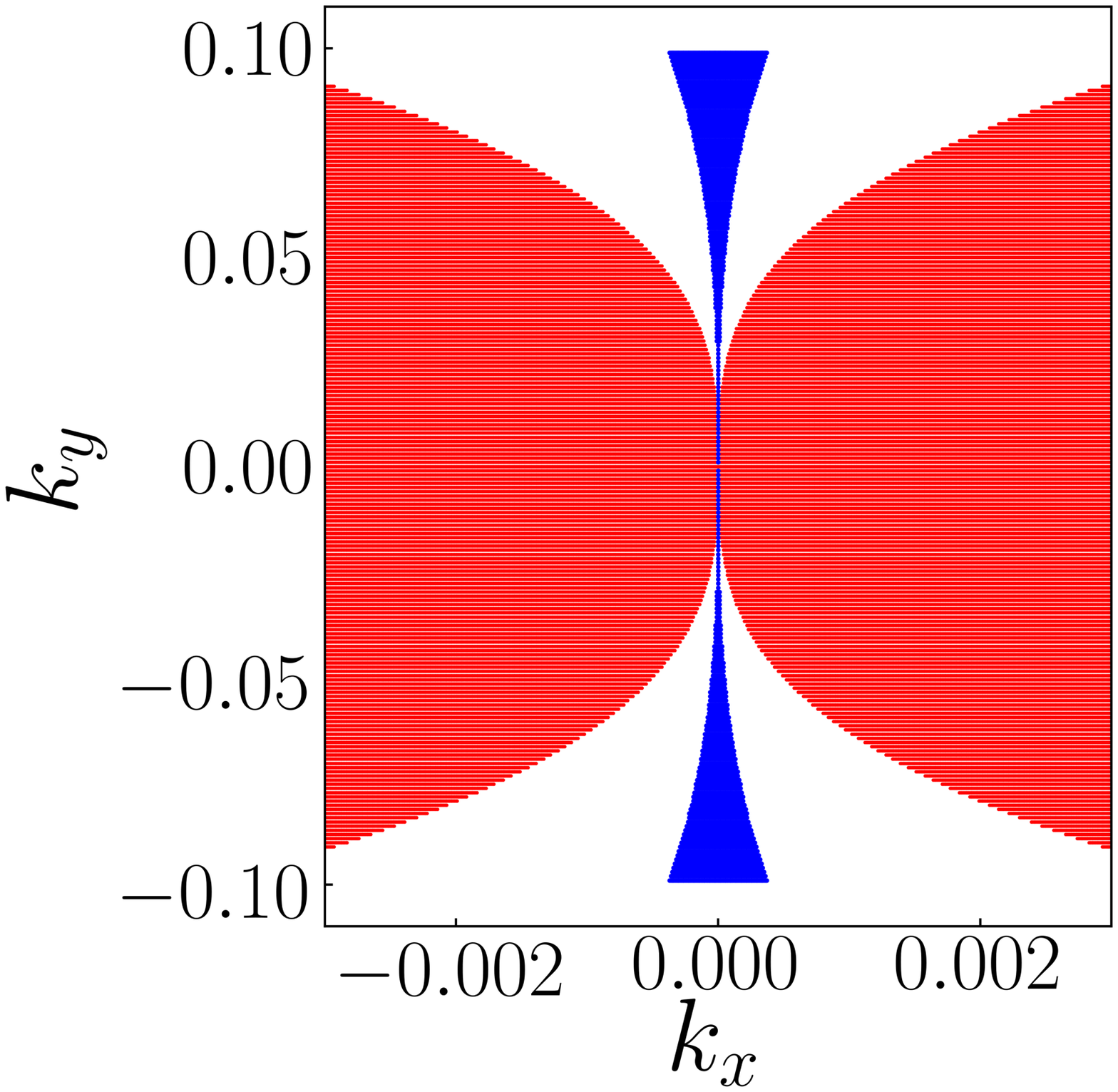} 
\vspace{-0.3cm}
\caption{(Color online) Schematic image of two regions (i) and (ii) for $\dot{\gamma}=\Gamma=1.0$, $\kappa=0.5$ and $r=0$. The blue and red regions, respectively, represent region (i) and (ii). The right side is the zoom of $k_x=0$.}
\vspace{-0.3cm}
\label{fig:dividing into two regions}
\end{figure}
In region (i), the dominant contribution of the $s$-integral in Eq.~(\ref{eq:correlation function: final expression}) comes from near
\begin{eqnarray}
s=\frac{1}{\Gamma (\kappa |\bm{k}|^2+r)},
\end{eqnarray}
where the integrand is approximated as
\begin{eqnarray}
e^{-\Gamma\big\{s(\kappa|\bm{k}|^2+r) + \frac{1}{2}\kappa\dot{\gamma} s^2k_xk_y + \frac{1}{12} \kappa\dot{\gamma}^2 s^3 k_x^2 \big\}} \approx e^{-\Gamma s(\kappa|\bm{k}|^2+r)}.
\end{eqnarray}
Then, Eq.~(\ref{eq:correlation function: final expression}) is approximately integrated as
\begin{eqnarray}
C_{\varphi\varphi}(\bm{k}) &\simeq& T\Gamma \int_0^{\infty} ds e^{-\Gamma s(\kappa |\bm{k}|^2+r) } \nonumber \\[3pt]
&=& \frac{T}{\kappa |\bm{k}|^2+r}.
\label{eq:linearized solution : region i}
\end{eqnarray}
Similarly, in region (ii), because the dominant contribution of the $s$-integral of Eq.~(\ref{eq:correlation function: final expression}) comes from near
\begin{eqnarray}
s = \Big(\frac{12}{\Gamma \kappa \dot{\gamma}^2 k_x^2}\Big)^{\frac{1}{3}},
\end{eqnarray}
$C_{\varphi\varphi}(\bm{k})$ is approximated as
\begin{eqnarray}
C_{\varphi\varphi}(\bm{k}) &\simeq& T\Gamma \int_0^{\infty} ds e^{- \frac{1}{12} \Gamma \kappa \dot{\gamma}^2 s^3 k_x^2 } \nonumber \\[3pt]
&=& \frac{12^{\frac{1}{3}}\Gamma(\frac{4}{3})T\Gamma^{\frac{2}{3}}}{\kappa^{\frac{1}{3}} \dot{\gamma}^{\frac{2}{3}} |k_x|^{\frac{2}{3}}} ,
\label{eq:linearized solution : region ii}
\end{eqnarray}
where $\Gamma(x)$ is the Gamma function.

We here notice that the behavior of Eqs.~(\ref{eq:linearized solution : region i}) and (\ref{eq:linearized solution : region ii}) is obtained from the limiting case of the following expression
\begin{eqnarray}
C_{\varphi\varphi}(\bm{k}) = \frac{T}{r + \kappa |\bm{k}|^2 + c_0(\Gamma^{-1} \sqrt{\kappa} \dot{\gamma} |k_x| )^{{\frac{2}{3}}}},
\end{eqnarray}
where $c_0=12^{\frac{1}{3}}\Gamma(\frac{4}{3})\simeq 2.04$. This expression describes well the behavior of $C_{\varphi\varphi}(\bm{k})$ in the colored regions of Fig.~\ref{fig:dividing into two regions}. From this expression, we find that the correlation length diverges as 
\begin{eqnarray}
\xi_x \sim r^{-3/2}  \ \ {\rm and} \ \ \xi_y \sim r^{-1/2}.
\end{eqnarray}

\subsection{Fluctuations around the ordered solution}
\label{subsec:Fluctuations in ordered state}
Next, we argue linear fluctuations around the ordered solution. For this purpose, we return to the model Eqs.~(1), (2), and (3) and consider the regime $r<0$. As explained in the main text, it is useful to decompose the field variable $\bm{\varphi}(\bm{r},t)$ into 
\begin{eqnarray}
\bm{\varphi}(\bm{r},t)=\Big(\sqrt{-\frac{r}{u}}+A(\bm{r},t)\Big) \big(\cos\theta(\bm{r},t), \sin\theta(\bm{r},t)\big).
\label{eq:decomposition of the field variable}
\end{eqnarray}
$A(\bm{r},t)$ and $\theta(\bm{r},t)$, respectively, correspond to the amplitude fluctuation and the phase fluctuation around the state realized at $T=0$. By substituting Eq.~(\ref{eq:decomposition of the field variable}) into Eqs.~(1), (2), and (3) and neglecting the non-linear terms, we obtain
\begin{eqnarray}
\Big[\frac{\partial}{\partial t} - \dot{\gamma} k_x \frac{\partial}{\partial k_y} +\Gamma (\kappa |\bm{k}|^2 +2|r|)\Big] \tilde{A} (\bm{k},t)= \tilde{\eta}_1(\bm{k},t),
\label{eq:phase mode: Fourier space}\\
\Big[\frac{\partial}{\partial t} - \dot{\gamma} k_x \frac{\partial}{\partial k_y} +\Gamma \kappa |\bm{k}|^2 \Big] \tilde{\theta} (\bm{k},t)= \tilde{\eta}_2(\bm{k},t).
\label{eq:phase mode: Fourier space}
\end{eqnarray}
Since Eq.~(\ref{eq:phase mode: Fourier space}) is equivalent to Eq.~(\ref{eq:explicit form of linearized order parameter dynamics in Fourier space}) with $r=0$, we can repeat the previous argument in Secs.~\ref{sec:Formal solution of linearized model} and \ref{sec:fluctuations near criticality}. The final expression is given by
\begin{eqnarray}
C_{\theta \theta}(\bm{k})  \simeq \frac{T}{c_0(\Gamma^{-1} \sqrt{\kappa} \dot{\gamma} k_x )^{{\frac{2}{3}}}+\kappa |\bm{k}|^2}.
\label{eq:correlation function: theta theta}
\end{eqnarray}
This equation is Eq.~(6) in the main text.

The amplitude correlation $C_{AA}(\bm{k})$, defined by $\langle \tilde{A}(\bm{k},t)\tilde{A}(\bm{k}',t)\rangle = C_{AA}(\bm{k})\delta(\bm{k}+\bm{k}')$, is also calculated as
\begin{eqnarray}
C_{AA}(\bm{k})  = T\Gamma \int_0^{\infty} ds e^{-\Gamma\big\{s(\kappa|\bm{k}|^2+2|r|) + \frac{1}{2}\kappa\dot{\gamma} s^2k_xk_y + \frac{1}{12} \kappa\dot{\gamma}^2 s^3 k_x^2 \big\}}.
\label{eq:correlation function: final expression: AA}
\end{eqnarray}
In the long-wavelength region, the main contribution of the $s$-integral arises from $s = 1/(2\Gamma |r|)$. Because 
\begin{eqnarray}
2\Gamma s |r| \gg \Gamma s \kappa|\bm{k}|^2, \  \frac{1}{2}\Gamma\kappa\dot{\gamma} s^2k_xk_y, \ \frac{1}{12} \Gamma \kappa\dot{\gamma}^2 s^3 k_x^2
\end{eqnarray}
holds near $s = 1/(2\Gamma |r|)$, Eq.~(\ref{eq:correlation function: final expression: AA}) is expanded as
\begin{eqnarray}
C_{AA}(\bm{k})  = T\Gamma \int_0^{\infty} ds e^{-2\Gamma s|r|}\Big(1 -\Gamma s \kappa|\bm{k}|^2-  \frac{\Gamma}{2}\kappa\dot{\gamma} s^2k_xk_y- \frac{\Gamma}{12} \kappa\dot{\gamma}^2 s^3 k_x^2 + \cdots \Big) .
\end{eqnarray}
Then, the $s$-integral is calculated as
\begin{eqnarray}
C_{AA}(\bm{k}) = \frac{T}{2|r|} - \frac{T\kappa}{4|r|^2}|\bm{k}|^2 - \frac{\dot{\gamma}}{\Gamma}\frac{T\kappa}{8|r|^3}k_x k_y \cdots.
\end{eqnarray}
Finally, to make it easier to see, we rewrite it in the Ornstein--Zernike form
\begin{eqnarray}
C_{AA}(\bm{k}) = \frac{T}{2|r| + \kappa |\bm{k}^2| + \dot{\gamma} \kappa k_xk_y/(2\Gamma|r|) + \cdots} .
\label{eq:correlation function: AA}
\end{eqnarray}

Eqs.~(\ref{eq:correlation function: theta theta}) and (\ref{eq:correlation function: AA}) give all the behaviors of fluctuations in the ordered state. In the real space, Eq.~(\ref{eq:correlation function: theta theta}) yields the power-law decay of the fluctuation, whereas Eq.~(\ref{eq:correlation function: AA}) gives the exponential decay. This result reflects the gapless nature of the phase fluctuation. We also find that the fractional exponent is specific to the phase fluctuations. The shear flow makes the amplitude fluctuation anisotropic without affecting the exponent.

\subsubsection*{Physical interpretation of suppression due to shear flow}
The shear flow induces stretching of fluctuations along the $x$-axis. This is the origin of anisotropy of the correlation function Eq.~(\ref{eq:correlation function: theta theta}). Here, let us physically estimate the decay rate originating from the stretching, which is denoted by $\lambda_s$. We fix the system size in the $y$-direction as $L_y$ and consider a fluctuation with $k_x$. Because the diffusion time in the $y$-direction is given by $L_y^2/\Gamma \kappa$, the stretching over the diffusion time is given by $\dot \gamma k_x L_y^3/(\Gamma \kappa)$. The decay rate $\lambda_s$ in the low temperature limit is then expressed as
\begin{equation}
  \lambda_s = \frac{\Gamma \kappa}{L_y^2} f\left(
  \frac{\dot \gamma k_x L_y^3}{\Gamma \kappa} \right),
\end{equation}  
where $f$ is a dimensionless function. Here, we reasonably assume that  $\lambda_s $ is independent of $L_y$ in the limit $\dot \gamma k_x L_y^3/(\Gamma \kappa) \to 0$. Then, the decay rate becomes
\begin{equation}
  \lambda_s \simeq
  \dot \gamma \left( k_x \sqrt{\frac{\Gamma \kappa}{\dot \gamma}} \right)^{2/3},
\end{equation}
which is so large that the fluctuation disappears faster than the diffusion process. As the result, the fluctuation of the order parameter along the $x$-axis is suppressed.

\section{Numerical implementation of uniform shear flow}
We explain a numerical implementation of shear flow.
The dynamics of the order parameter is given by Eqs.~(1), (2), and (3), which are explicitly written as
\begin{eqnarray}
\frac{\partial \varphi_a}{\partial t}  + \dot{\gamma} y \frac{\partial \varphi_a}{\partial x} &=& - \Gamma \Big(-\kappa \Delta + r + u |\bm{\varphi}|^2 \Big) \varphi_a+ \eta_a.
\label{eq:explicit form of order parameter dynamics}
\end{eqnarray}

Numerically solving Eq.~(\ref{eq:explicit form of order parameter dynamics}) in the Cartecian coordinate system is difficult because the term $\dot{\gamma} y \frac{\partial \varphi_a}{\partial x} $ at $y=L_y$ becomes larger in proportion to $L_y$. This difficulty is removed by using a new coordinates system moving with velocity $-\bm{v}$, which was firstly proposed by Toh \textit{et al.}~\cite{smToh1991} and secondly by Onuki~\cite{smOnuki1997}. Here, we briefly review this method.

First, we introduce the new coordinate system $(\bm{r}',t')=(x',y',t')$ defined by
\begin{eqnarray}
x' = x - \dot{\gamma} t y \ , \ y'=y \ , \ t'=t.
\end{eqnarray}
The order parameter in this coordinate system is given by $\hat{\varphi}_a(\bm{r}',t') = \varphi_a(\bm{r},t)$. The dynamics of $\hat{\varphi}_a(\bm{r}',t')$ is then derived from Eq.~(\ref{eq:explicit form of order parameter dynamics}) as
\begin{eqnarray}
\frac{\partial \hat{\varphi}_a}{\partial t'}  &=& - \Gamma \Big(-\kappa \Delta' + r + u |\hat{\bm{\varphi}}|^2 \Big) \hat{\varphi}_a+ \hat{\eta}_a
\label{eq:explicit form of order parameter dynamics: new coordinate systems}
\end{eqnarray}
with
\begin{eqnarray}
\Delta' = \Big(\frac{\partial}{\partial x'}\Big)^2 + \Big(\frac{\partial}{\partial y'} - \dot{\gamma} t \frac{\partial}{\partial x'} \Big)^2.
\end{eqnarray}

Eq.~(\ref{eq:explicit form of order parameter dynamics: new coordinate systems}) does not contain the term proportional to $y$, but instead contains the term proportional to $t$. For $0\leq t \leq 1/\dot{\gamma}$, this term is numerically stable because it remains $O(1)$. However, this term becomes so large for $t\gg 1/\dot{\gamma}$. In order to overcome this difficulty, we repeat the coordinate transformation at every $1/\dot{\gamma}$. This procedure is summarized as follows.
\begin{enumerate}
\item Transform $\varphi_a(\bm{r},t)$ into $\hat{\varphi}_a(\bm{r}',t')$ at $t=0$.
\item Solve Eq.~(\ref{eq:explicit form of order parameter dynamics: new coordinate systems}) until $t'=1/\dot{\gamma}$ with the initial condition $\hat{\varphi}_a(\bm{r}',t'=0)$.
\item Transform $\hat{\varphi}_a(\bm{r}',t')$ into $\varphi_a(\bm{r},t)$ at $t'=1/\dot{\gamma}$.
\item Reset $t'$ from $1/\dot{\gamma}$ to $0$, and start again from procedure 1.
\end{enumerate}

In the numerical simulations, Eq.~(\ref{eq:explicit form of order parameter dynamics: new coordinate systems}) is discretized with the time step $\delta t = 0.01$ and space interval $\delta x=1.0$. The time integration is performed via the optimal stochastic Runge--Kutta scheme of order (2,2) in Ref.~\cite{smDebrabant2008}. We impose the standard periodic boundary condition along the $x$-axis and the Lees-Edwards periodic boundary condition along the $y$-axis~\cite{smLees1972}.

\section{Details of finite-size scaling theory}
As mentioned in the main text, the finite-size scaling theory is constructed on the two assumptions:
\begin{assumption}
A free energy $F(\tau,h,L^{-1}_x,L^{-1}_y;\dot{\gamma})$ for the finite-size system satisfies the scaling relation near the critical point
\begin{eqnarray}
F(\tau,h,L^{-1}_x,L^{-1}_y;\dot{\gamma}) = F(b^{z_\tau}\tau,b^{z_h}h,b^{z_x}L^{-1}_x,bL^{-1}_y;\dot{\gamma}).
\label{eq:assumption 1: appendix}
\end{eqnarray}
Here, $\tau=(r-r_c)/r_c$ is the dimensionless distance from the transition point $r_c$, and $h$ is the external field that gives
\begin{eqnarray}
\langle \hat{m}\rangle &=& -\frac{1}{L_x  L_y}\frac{\partial F}{\partial h}, \\
\chi &=& L_x L_y\big(\langle \hat{m}^2\rangle - \langle \hat{m}\rangle^2\big) = -\frac{1}{L_x  L_y} \frac{\partial^2 F}{\partial h^2}.
\label{eq:def of chi: appendix}
\end{eqnarray}
\end{assumption}
\begin{assumption}
Near the critical point, any quantity $\langle \hat{A}\rangle_{h=0}$ can be expressed in terms of two correlation lengths $\xi_x$ and $\xi_y$ as
\begin{eqnarray}
\langle \hat{A} \rangle_{h=0}(L^{-1}_x,L^{-1}_y,\tau;\dot{\gamma}) = L_x^{w_A}\mathcal{A}\Big(\frac{\xi_x}{L_x},\frac{\xi_y}{L_y};\dot{\gamma} \Big),
\end{eqnarray}
where $w_A$ is a constant independent of $\dot{\gamma}$, and $\mathcal{A}$ is a scaling function.
\end{assumption}
Below, we summarize the important results derived from these assumptions.

\subsection{System-size dependence of the various quantities}
The system-size dependence of various quantities such as Eqs.~(9) and (10) are calculated from Assumption 1. Differentiating Eq.~(\ref{eq:assumption 1: appendix}) with the scaling field $h$ and substituting $h=0$ lead to 
\begin{eqnarray}
\langle \hat{m}\rangle_{h=0}(\tau,L_x^{-1},L_y^{-1};\dot{\gamma}) =  b^{-(1+z_x)+z_h}\langle \hat{m}\rangle_{h=0}(b^{z_\tau}\tau,b^{z_x}L_x^{-1},bL_y^{-1};\dot{\gamma}).
\label{eq:scaling relation of magnetization : intermediate 1}
\end{eqnarray}
Because $b$ can be arbitrarily chosen, especially by substituting $b=L_y$ into Eq.~(\ref{eq:scaling relation of magnetization : intermediate 1}), we obtain
\begin{eqnarray}
\langle \hat{m}\rangle_{h=0}(\tau,L_x^{-1},L_y^{-1};\dot{\gamma}) =  L_y^{\tilde{z}_h}\langle \hat{m}\rangle_{h=0}(L_y^{z_\tau}\tau,L_y^{z_x}L_x^{-1},1;\dot{\gamma}),
\label{eq:scaling relation of magnetization : intermediate 2}
\end{eqnarray}
where we have introduced $\tilde{z}_h\equiv -(1+z_x) + z_h$ for later convenience.
In the similar way, we find that the $n$-th moment of magnetization satisfies
\begin{eqnarray}
\langle \hat{m}^n\rangle_{h=0}(\tau,L_x^{-1},L_y^{-1};\dot{\gamma}) =  L_y^{\tilde{z}_h n}\langle \hat{m}^n\rangle_{h=0}(L_y^{z_\tau}\tau,L_y^{z_x}L_x^{-1},1;\dot{\gamma}).
\label{eq:scaling relation of k-th moment : intermediate 1}
\end{eqnarray}
Combining Eq.~(\ref{eq:scaling relation of k-th moment : intermediate 1}) with the Binder parameter defined by
\begin{eqnarray}
U(\tau,L_x^{-1},L_y^{-1};\dot{\gamma}) \equiv \frac{\langle \hat{m}^4\rangle_{h=0}}{\big(\langle \hat{m}^2\rangle_{h=0}\big)^2},
\end{eqnarray}
we have the relation
\begin{eqnarray}
U(\tau,L_x^{-1},L_y^{-1};\dot{\gamma}) = \mathcal{U}(L_y^{z_\tau}\tau,L_y^{z_x}L_x^{-1};\dot{\gamma}),
\label{eq:finite-size scaling behavior of binder parameter}
\end{eqnarray}
where $\mathcal{U}$ is a scaling function.

\subsection{Expression of the critical exponent}
All the critical exponents are expressed by combining the scaling exponents $z_\tau$, $z_h$, and $z_x$. First, we consider the critical exponents $\nu_x$ and $\nu_y$ that characterize the divergence of the correlation length at the critical point:
\begin{eqnarray}
\xi_x \sim |\tau|^{-\nu_x} \ {\rm and} \ \ \xi_y \sim |\tau|^{-\nu_y} .
\end{eqnarray}
By setting $b=\tau^{-1/z_\tau}$ in Eq.~(\ref{eq:assumption 1: appendix}), we have
\begin{eqnarray}
F(\tau,h,L_x^{-1},L_y^{-1};\dot{\gamma}) = F(1,\tau^{-z_h/z_\tau} h, \tau^{-z_x/z_\tau}L_x^{-1},\tau^{-1/z_\tau}L_y^{-1};\dot{\gamma}).
\label{eq:expression of critical exponent: intermediate 1}
\end{eqnarray}
Then, by applying Assumption 2 to the right-hand side of Eq.~(\ref{eq:expression of critical exponent: intermediate 1}), the free energy is expressed as
\begin{eqnarray}
F(1,h=0, \tau^{-z_x/z_\tau}L_x^{-1},\tau^{-1/z_\tau}L_y^{-1};\dot{\gamma}) = L_x^{\omega_f} \mathcal{F}\Big(\frac{\xi_x}{L_x},\frac{\xi_y}{L_y};\dot{\gamma} \Big),
\end{eqnarray}
where $\mathcal{F}$ is a scaling function.
From the comparison of both sides of this equation, we find that $\omega_f$ is equal to $0$, and $\xi_x$ and $\xi_y$ are related with $\tau$ as
\begin{eqnarray}
\xi_x \sim \tau^{-z_x/z_\tau} \ {\rm and} \ \ \xi_y \sim \tau^{-1/z_\tau}.
\label{eq:relationship between xi and t}
\end{eqnarray}
Accordingly, the critical exponents $\nu_x$ and $\nu_y$ turn out to be written as
\begin{eqnarray}
\nu_x = \frac{z_x}{z_\tau} \ {\rm and} \ \ \nu_y = \frac{1}{z_\tau}.
\label{eq:expression of nu}
\end{eqnarray}

Next, we consider the critical exponent $\beta$ that characterizes the onset of the magnetization slightly below the critical point:
\begin{eqnarray}
m \sim |\tau|^{\beta}.
\end{eqnarray}
In order to express $\beta$ with $(z_\tau,z_h,z_x)$, we return to Eq.~(\ref{eq:scaling relation of magnetization : intermediate 1}). Substituting $b=|\tau|^{-1/z_\tau}$ into Eq.~(\ref{eq:scaling relation of magnetization : intermediate 1}), we have
\begin{eqnarray}
\langle \hat{m}\rangle_{h=0}(\tau,L_x^{-1},L_y^{-1};\dot{\gamma}) =  |\tau|^{-\tilde{z}_h/z_\tau}\langle \hat{m}\rangle_{h=0}(1,|\tau|^{-z_x/z_\tau}L_x^{-1},|\tau|^{-1/z_\tau}L_y^{-1};\dot{\gamma}).
\label{eq:scaling relation of magnetization : intermediate 3}
\end{eqnarray}
Furthermore, by using Eq.~(\ref{eq:relationship between xi and t}), Eq.~(\ref{eq:scaling relation of magnetization : intermediate 3}) is rewritten as
\begin{eqnarray}
\langle \hat{m}\rangle_{h=0}(\tau,L_x^{-1},L_y^{-1};\dot{\gamma}) = |\tau|^{-\tilde{z}_h/z_\tau}\mathcal{M}\Big(\frac{\xi_x}{L_x},\frac{\xi_y}{L_y};\dot{\gamma} \Big),
\end{eqnarray}
where $\mathcal{M}$ is an appropriate scaling function. Then, the onset of magnetization in the infinite system is given by
\begin{eqnarray}
\langle \hat{m}\rangle_{h=0}(\tau,L_x^{-1}\to0,L_y^{-1}\to0;\dot{\gamma}) \sim |\tau|^{-\tilde{z}_h/z_\tau},
\end{eqnarray}
which leads to
\begin{eqnarray}
\beta = -\frac{\tilde{z}_h}{z_\tau}.
\label{eq:expression of beta}
\end{eqnarray}

Finally, we consider the critical exponent $\gamma$ that characterizes the singularity of the susceptibility at the critical point:
\begin{eqnarray}
\chi \sim |\tau|^{-\gamma}.
\end{eqnarray}
We start with the second-order derivative of Eq.~(\ref{eq:assumption 1: appendix}). By noting that it is related to $\chi$ through Eq.~(\ref{eq:def of chi: appendix}), we have
\begin{eqnarray}
\chi(\tau,h=0,L_x^{-1},L_y^{-1};\dot{\gamma}) =  b^{2\tilde{z}_h + z_x+ 1}\chi(b^{z_\tau}\tau,h=0,b^{z_{Lx}}L_x^{-1},bL_y^{-1}).
\label{eq: scaling relation of chi: intermediate 1}
\end{eqnarray}
By setting $b=|\tau|^{-1/z_\tau}$ and using Eq.~(\ref{eq:relationship between xi and t}), Eq.~(\ref{eq: scaling relation of chi: intermediate 1}) is rewritten as
\begin{eqnarray}
\chi(\tau,h=0,L_x^{-1},L_y^{-1};\dot{\gamma}) =   |\tau|^{-(2\tilde{z}_h + z_x+ 1)/z_\tau}\mathcal{X}\Big(\frac{\xi_x}{L_x},\frac{\xi_y}{L_y};\dot{\gamma} \Big),
\end{eqnarray}
where $\mathcal{X}$ is a scaling function. Accordingly, $\gamma$ is given by
\begin{eqnarray}
\gamma = \frac{2\tilde{z}_h + z_x+ 1}{z_\tau}.
\label{eq:expression of gamma}
\end{eqnarray}

It is worthwhile to note that there are only three independent critical exponents because all the critical exponents are expressed by combination of three scaling exponents $z_\tau$, $z_h$, and $z_x$. In other words, the four critical exponents $\nu_x$, $\nu_y$, $\beta$, and $\gamma$ are not independent. Actually, from Eqs.~(\ref{eq:expression of nu}), (\ref{eq:expression of beta}) and (\ref{eq:expression of gamma}), we derive the hyperscaling relation
\begin{eqnarray}
2\beta+\gamma = \nu_x+\nu_y.
\label{eq:hyper scaling relation: appendix}
\end{eqnarray}
It is well known that this hyperscaling relation holds in anisotropic systems~\cite{smBinder1989,smWang1996,smAlbano2002,smWinter2010,smHucht2012}.

\section{Supplemental simulation results}
We provide the supplemental simulation results for the completeness of this work. In all simulations, we take the ensemble average over $32$ noise realizations and the time average over $10^6$ different times at $t=100i \delta t$.

\subsection{Finite-size scaling analysis for $\dot{\gamma}=0.01$ and $\dot{\gamma}=0.5$}
We present the simulation data of the Binder parameter $U$ for $\dot{\gamma}=0.01$ and $\dot{\gamma}=0.5$ in Fig.~\ref{fig:plot of binder parameter_smaller}. We use different system sizes as $(L_x,L_y) = (125,100), (216,120), {\rm and} \ (343,140)$, which satisfy $L_y = 20L_x^{1/3}$ (i.e. $z_x$ is fixed at $3$). These figures indicate the existence of the unique intersection point as for the case $\dot{\gamma}=5.0$ in the main text. 
\begin{figure}[t]
\centering
\includegraphics[width=8cm]{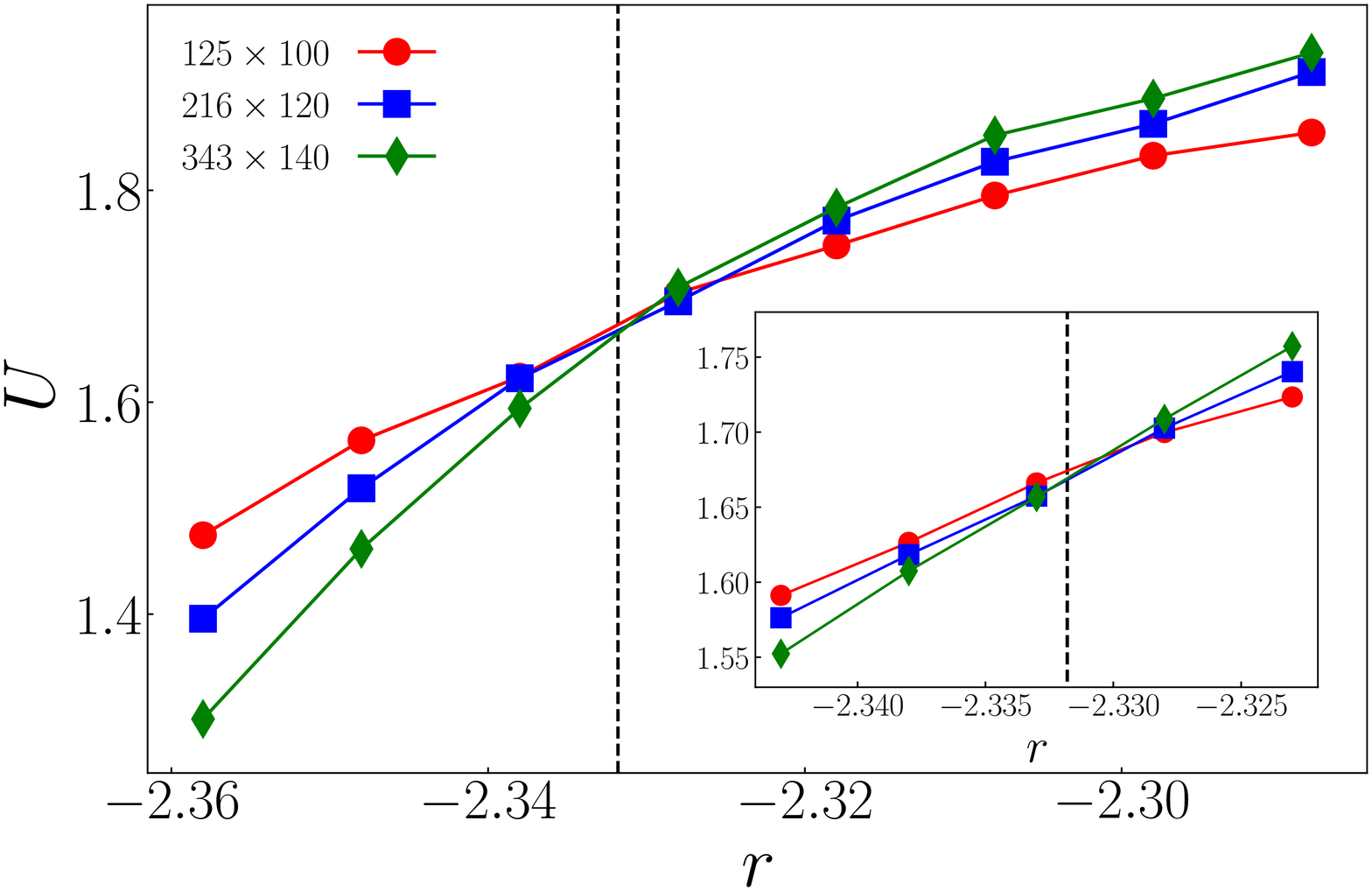} 
\includegraphics[width=8cm]{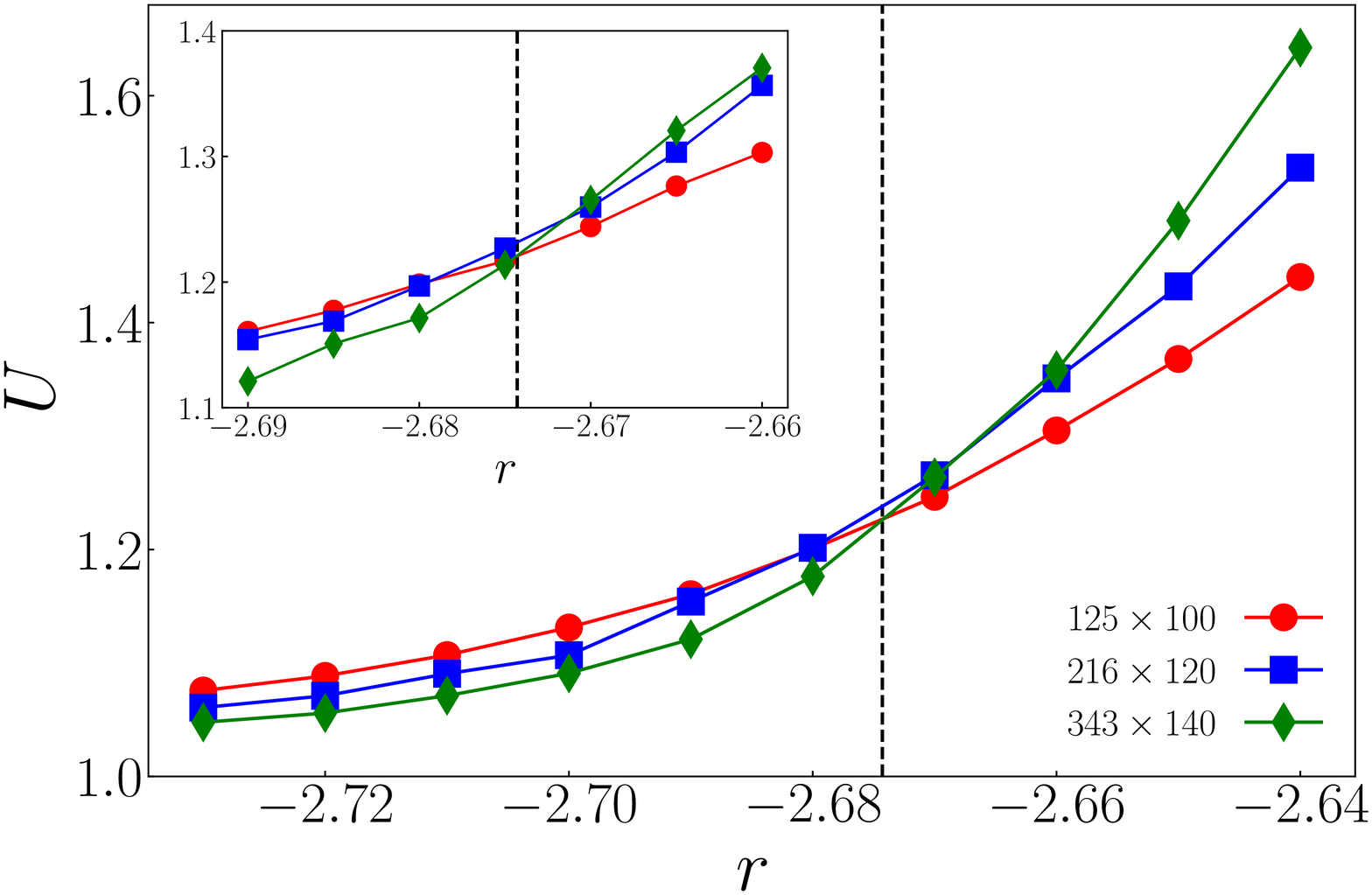}
\vspace{-0.3cm}
\caption{(Color online) Same as Fig.~1, but with $\dot{\gamma}=0.5$ (left) and $\dot{\gamma}=0.01$ (right). The error bars are in the order of the point size.}
\vspace{-0.3cm}
\label{fig:plot of binder parameter_smaller}
\end{figure}
The critical exponents are estimated by simultaneously fitting the magnetization $\langle \hat{m}\rangle$ and the Binder parameter $U$ to Eqs.~(11) and (12). The result is summarized in Tab.~\ref{tab:summary of scaling dimension}. Furthermore, we test another estimation method as a consistency check; See the next subsection for details. We list the result in Tab.~\ref{tab:summary of scaling dimension}.
\begin{table}[b]
\begin{center}
\begin{tabularx}{170mm}{C||CCC|CCC}\hline
& \multicolumn{3}{c|}{Fitting} & \multicolumn{3}{c}{At criticality}\\
$\dot{\gamma}$ & $\tilde{z}_h$ & $z_\tau$ & $\beta$ & $\tilde{z}_h$ & $z_\tau$ & $\beta$ \\ \hline\hline
$5.0$ & $-0.980 \pm0.026$ & $2.05\pm0.11$ & $0.480\pm0.029$ & $-0.990\pm0.007$ & $1.98\pm 0.01$ & $0.500\pm0.004$ \\
$0.5$ & $-0.989 \pm 0.028$ & $1.83\pm0.13$ & $0.540\pm0.041$ &  $-0.941\pm0.004$ & $1.85\pm 0.13$ & $0.509 \pm 0.036$ \\
$0.01$ & $-0.579\pm0.020$ & $2.03\pm0.20$ & $0.285\pm0.030$ & $-0.566\pm0.004$ & $2.14\pm0.12$ & $0.264\pm0.015$ \\\hline\hline
\end{tabularx}
\caption{Scaling exponents estimated from the numerical simulation. ``Fitting" means the value obtained by fitting the simulation data of $\langle \hat{m}\rangle_{h=0}$ and $U$ to Eqs.~(11) and (12). ``At criticality" means the one obtained from the data of magnetization at the critical point by using the method explained in Sec.~\ref{subsec: data at critical point}.}
\label{tab:summary of scaling dimension}
\end{center}
\vspace{-0.7cm}
\end{table}

These two estimations are consistent with each other, revealing the validity of our estimation method. We find that the critical exponent $\beta$ for $\dot{\gamma}=5.0$ and $0.5$ is accurately characterized by the mean-field theory. However, for $\dot{\gamma}=0.01$, there is relatively large deviation between the estimation result and the mean-field theory. To make it more clear, we plot the scaled magnetization data in Fig.~\ref{fig:plot of rescaled magnetization}. $\tilde{z}_h$ and $z_\tau$ are fixed at the mean-field value, $\tilde{z}_h=-1$ and $z_\tau=2$. For $\dot{\gamma}=0.5$, we find that the scaled data for the different system size are superimposed on a single curve. It means that the finite-size scaling relation Eq.~(11) holds with $\tilde{z}_h=-1$ and $z_\tau=2$. In contrast, for $\dot{\gamma}=0.01$ the scaled data do not overlap as expected. Therefore, the mean-field theory may not be applicable to the case with small shear rate.
\begin{figure}[t]
\centering
\includegraphics[width=8cm]{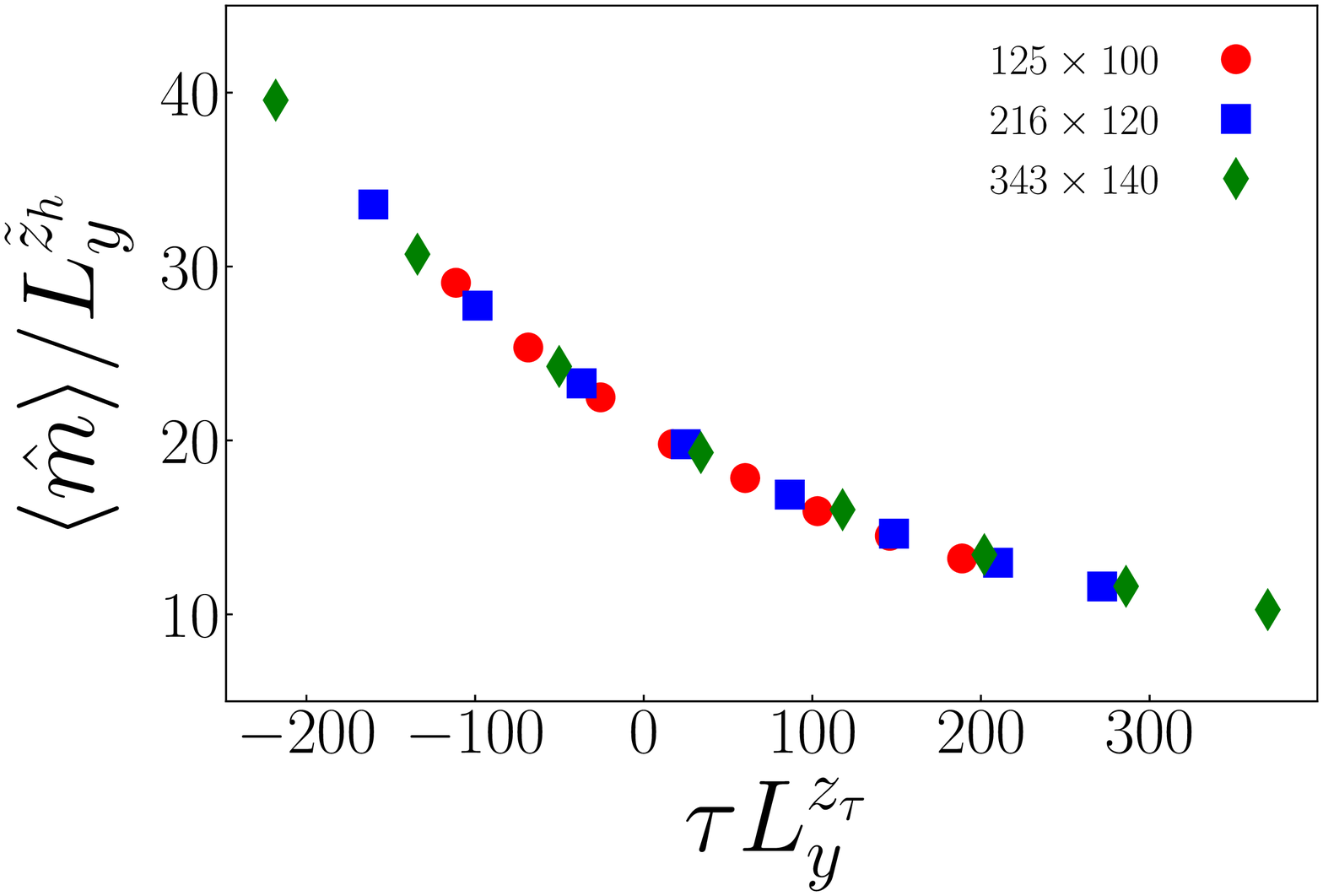} 
\includegraphics[width=8cm]{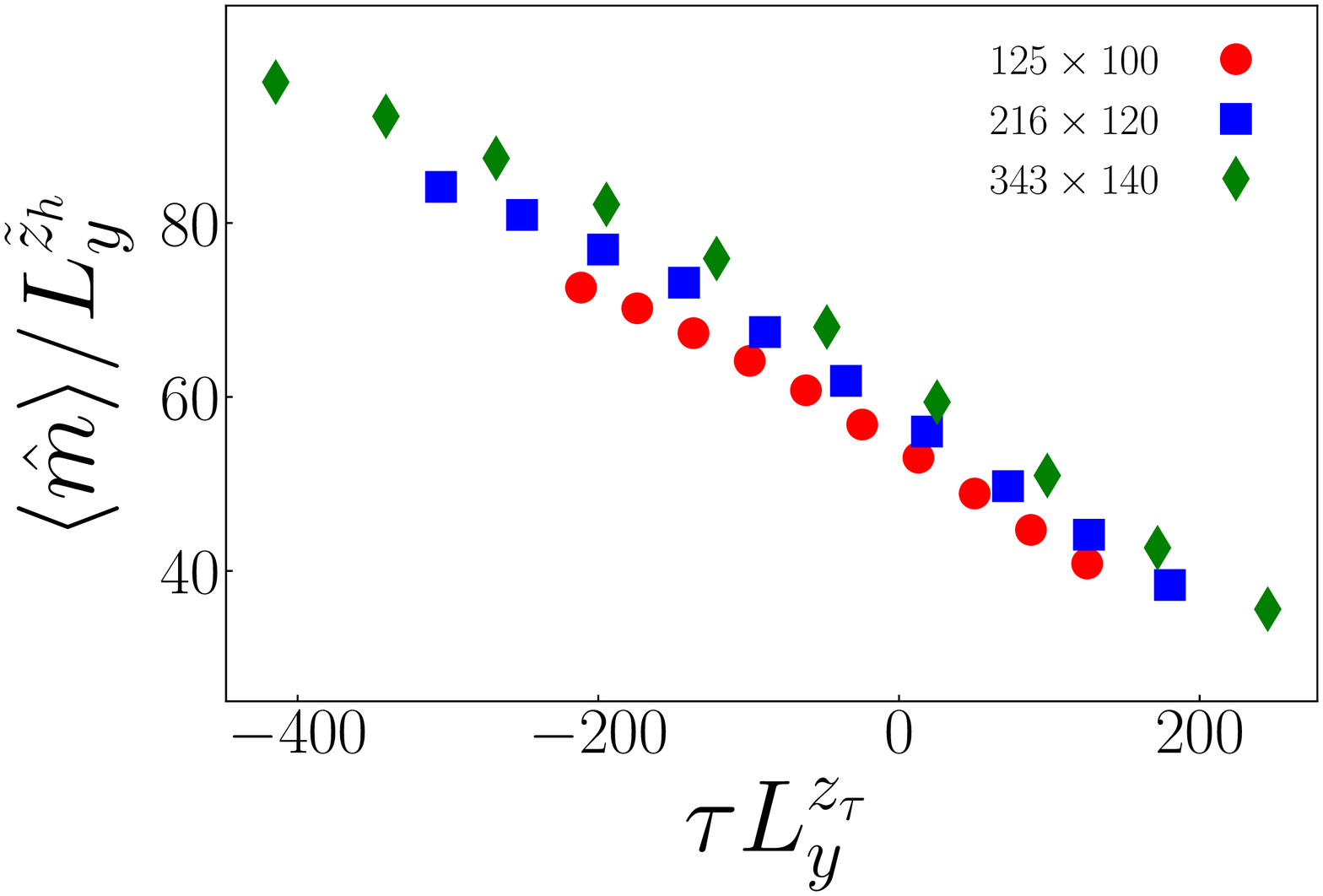}
\vspace{-0.3cm}
\caption{(Color online) The finite-size scaling plot of magnetization. Left: $\dot{\gamma}=0.5$, Right $\dot{\gamma}=0.01$. $\tilde{z}_h$ and $z_\tau$ are fixed at the mean-field value, $\tilde{z}_h=-1$ and $z_\tau=2$. The error bars are in the order of the point size.}
\vspace{-0.3cm}
\label{fig:plot of rescaled magnetization}
\end{figure}

\subsection{Data at the critical point}
\label{subsec: data at critical point}
As a consistency check, we conduct another estimation method that uses the data at the critical point. Here, we explain the method and the result.

According to the finite-size scaling relation Eq.~(11), the magnetization behaves as 
\begin{eqnarray}
\langle \hat{m}\rangle_{h=0} = L_y^{\tilde{z}_h} \mathcal{M}_1(\tau=0,L_y^{z_x}L_x^{-1};\dot{\gamma}), \\
\frac{d\langle \hat{m}\rangle_{h=0}}{d\tau} = L_y^{\tilde{z}_h+z_\tau} \mathcal{M}'_1(\tau=0,L_y^{z_x}L_x^{-1};\dot{\gamma}) 
\end{eqnarray}
right at the critical point. Therefore, when the system size increases with $L_y^{z_x}L_x^{-1}$ fixed, the simulation data for $\langle \hat{m}\rangle_{h=0}$ and $d\langle \hat{m}\rangle_{h=0}/d\tau$ are fitted by the simple relations
\begin{eqnarray}
\langle \hat{m}\rangle_{h=0} = a  L_y^{\tilde{z}_h} ,
\label{eq:simple relations for magnetization scaling}\\
\frac{d\langle \hat{m}\rangle_{h=0}}{d\tau} = b L_y^{\tilde{z}_h+z_\tau}.
\label{eq:simple relations for magnetization derivative scaling}
\end{eqnarray}
Furthermore, from the similar argument, we have the relation for the Binder parameter at the critical point:
\begin{eqnarray}
\frac{dU}{d\tau} = c L_y^{z_\tau}.
\label{eq:simple relations for Binderparameter derivative scaling}
\end{eqnarray}
In Fig.~\ref{fig:plot of scaling of magnetization}, we plot them for $\dot{\gamma}=5.0$, $0.5$, and $0.01$ with a log-log plot. From these figures, we find that all the data can be well described by (\ref{eq:simple relations for magnetization scaling}), (\ref{eq:simple relations for magnetization derivative scaling}), and (\ref{eq:simple relations for Binderparameter derivative scaling}) as expected from the finite-size scaling theory. 

Once we confirm the ansatz of the finite scaling theory, the exponent $\tilde{z}_h$ is estimated from the slope of $\langle\hat{m}\rangle_{h=0}$, and the exponent $z_{\tau}$ is estimated from the slope of $d\langle\hat{m}\rangle_{h=0}/d\tau$ or that of $dU/d\tau$. The two estimations of $z_{\tau}$ are in good agreement with each other. We especially choose the exponents obtained from the magnetization and summarize them in Tab.~\ref{tab:summary of scaling dimension}. As explained above, the exponents obtained from the data at criticality are consistent with that obtained by fitting the data over a wide region to Eqs.~(11) and (12). This result increases the validity of our analysis.
\begin{figure}[t]
\centering
\subfigure[$r=-1.9260$ ($r_c=-1.9257$), $\dot{\gamma}=5.0$]{
\includegraphics[width=6cm]{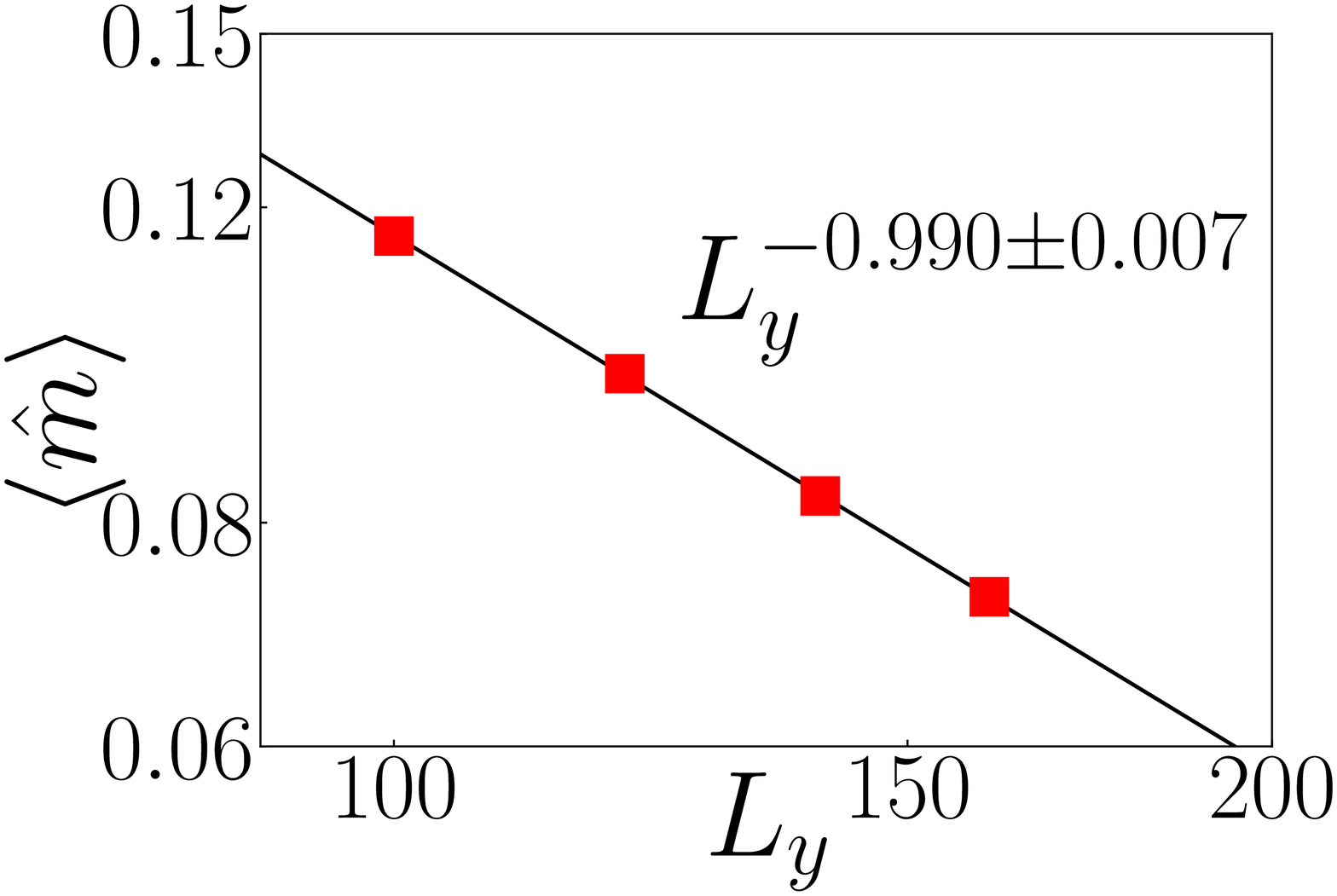} 
\includegraphics[width=6cm]{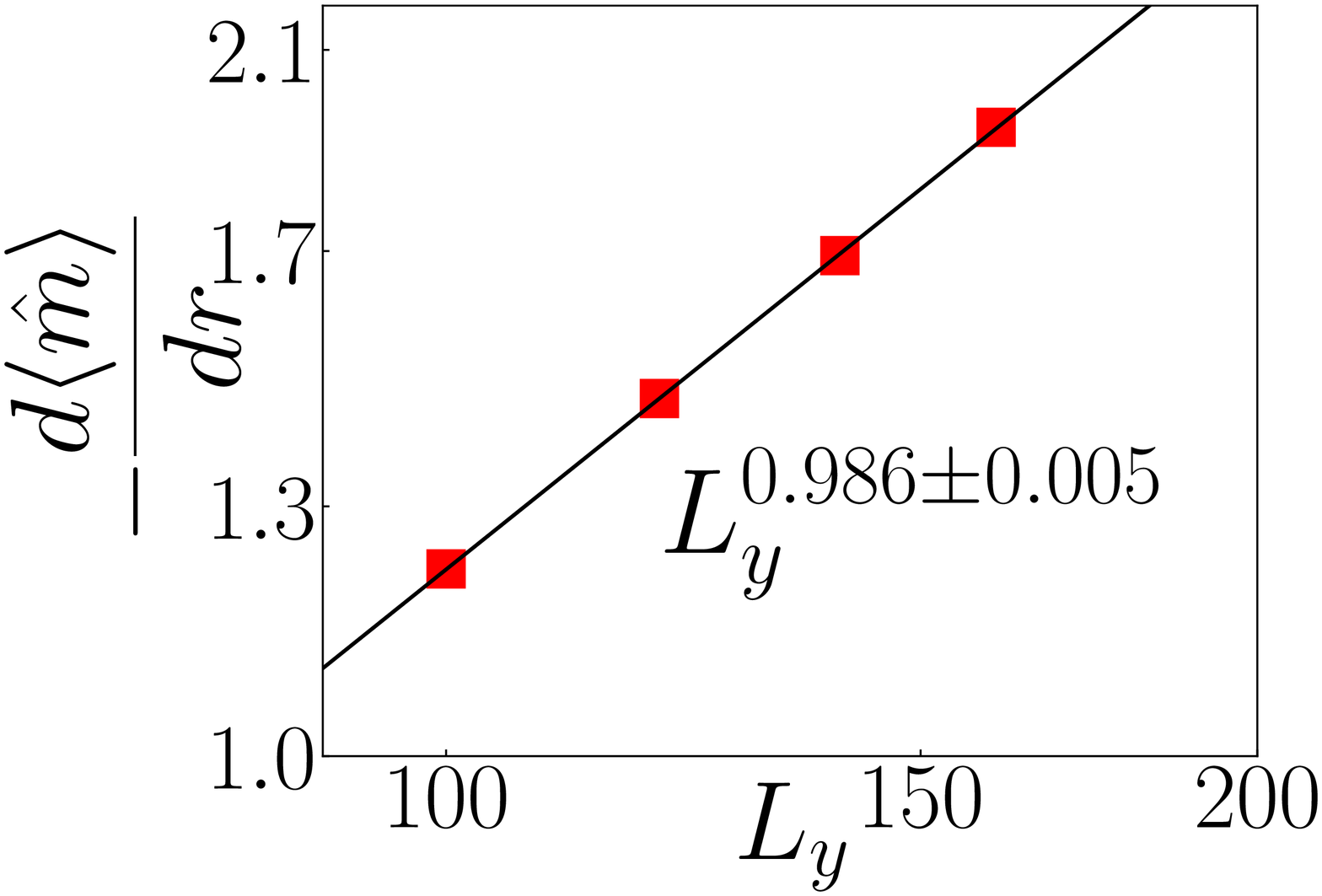}
\includegraphics[width=6cm]{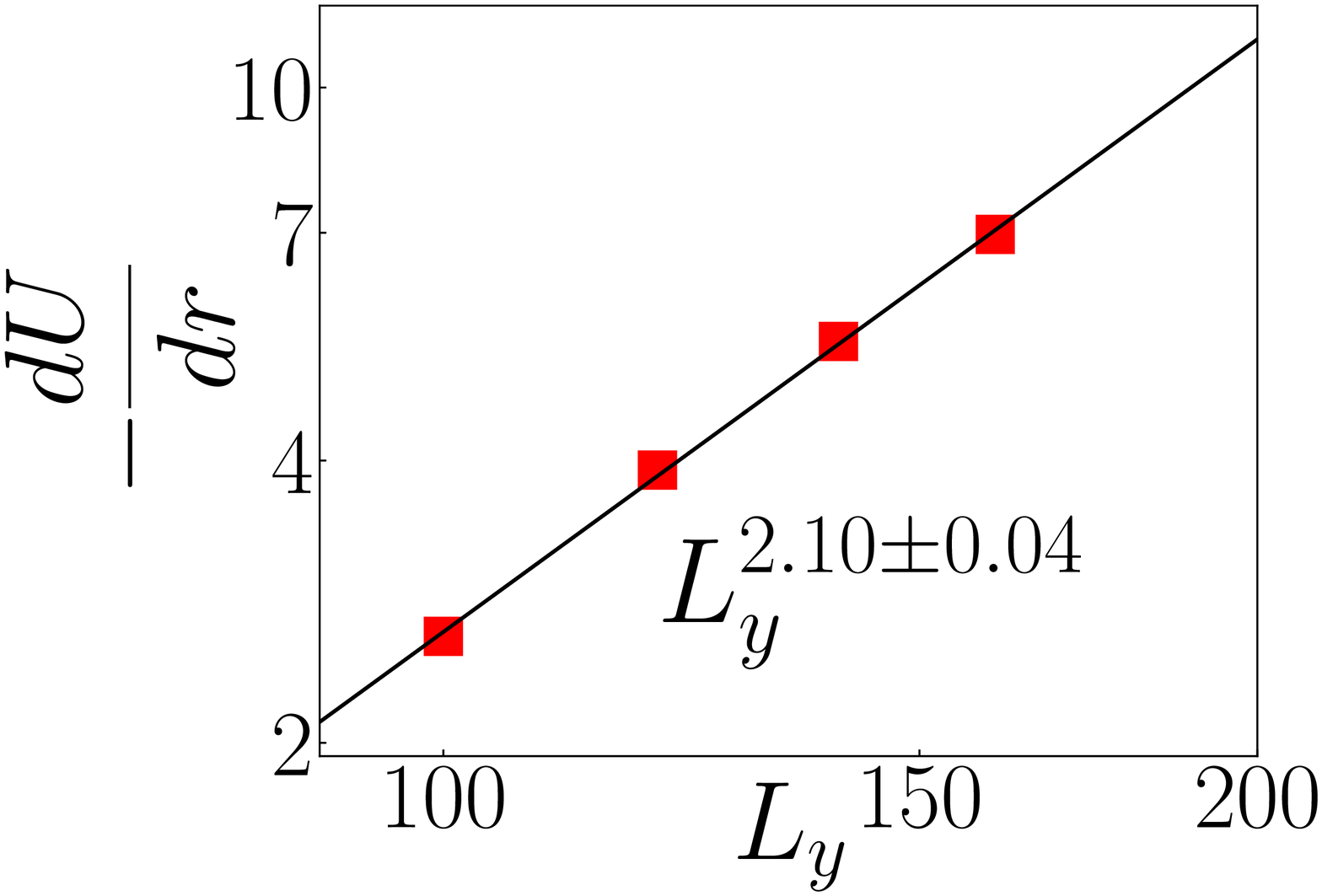}
}\\
\subfigure[$r=-2.3325$ ($r_c=-2.3318$), $\dot{\gamma}=0.5$]{
\includegraphics[width=6cm]{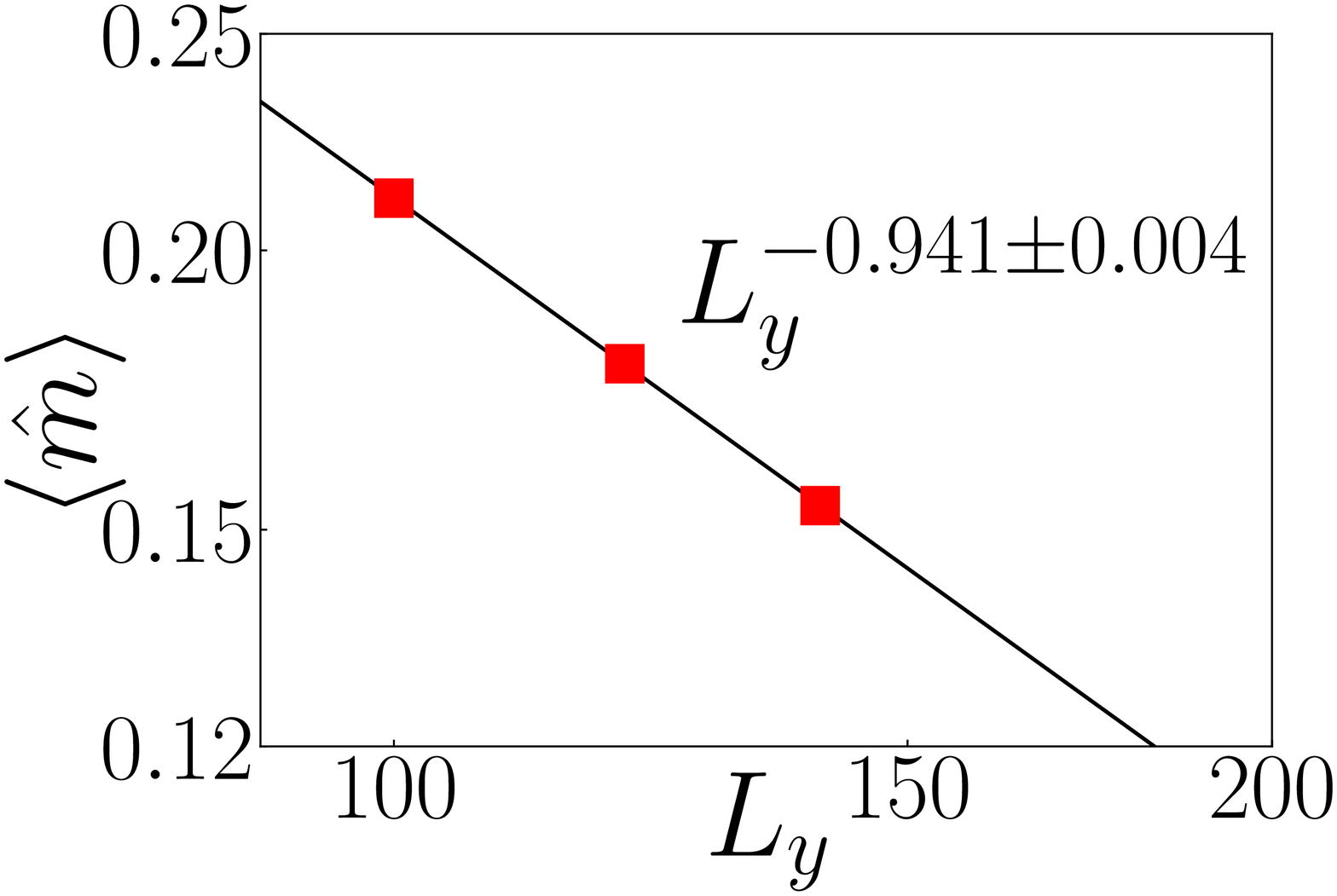} 
\includegraphics[width=6cm]{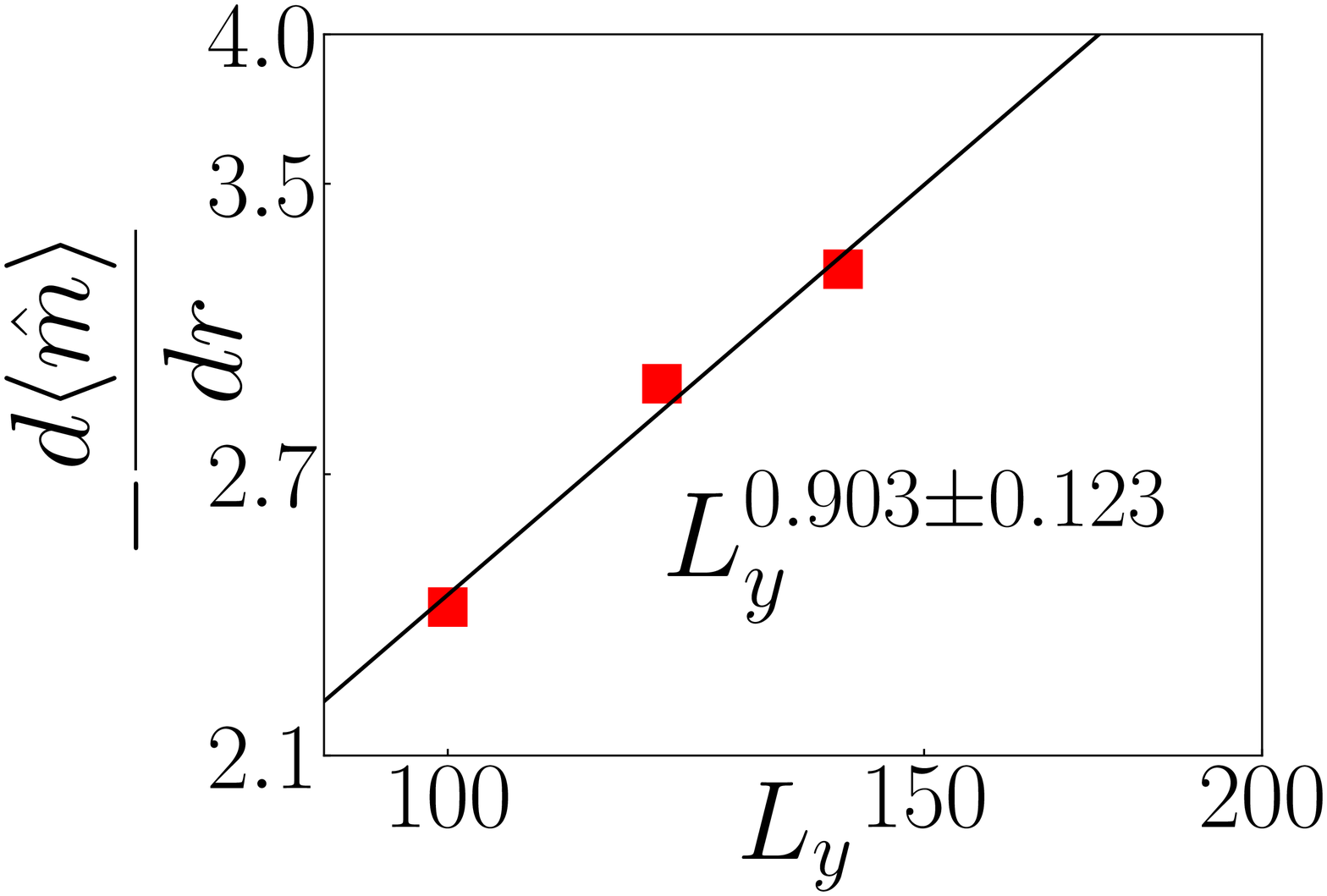}
\includegraphics[width=6cm]{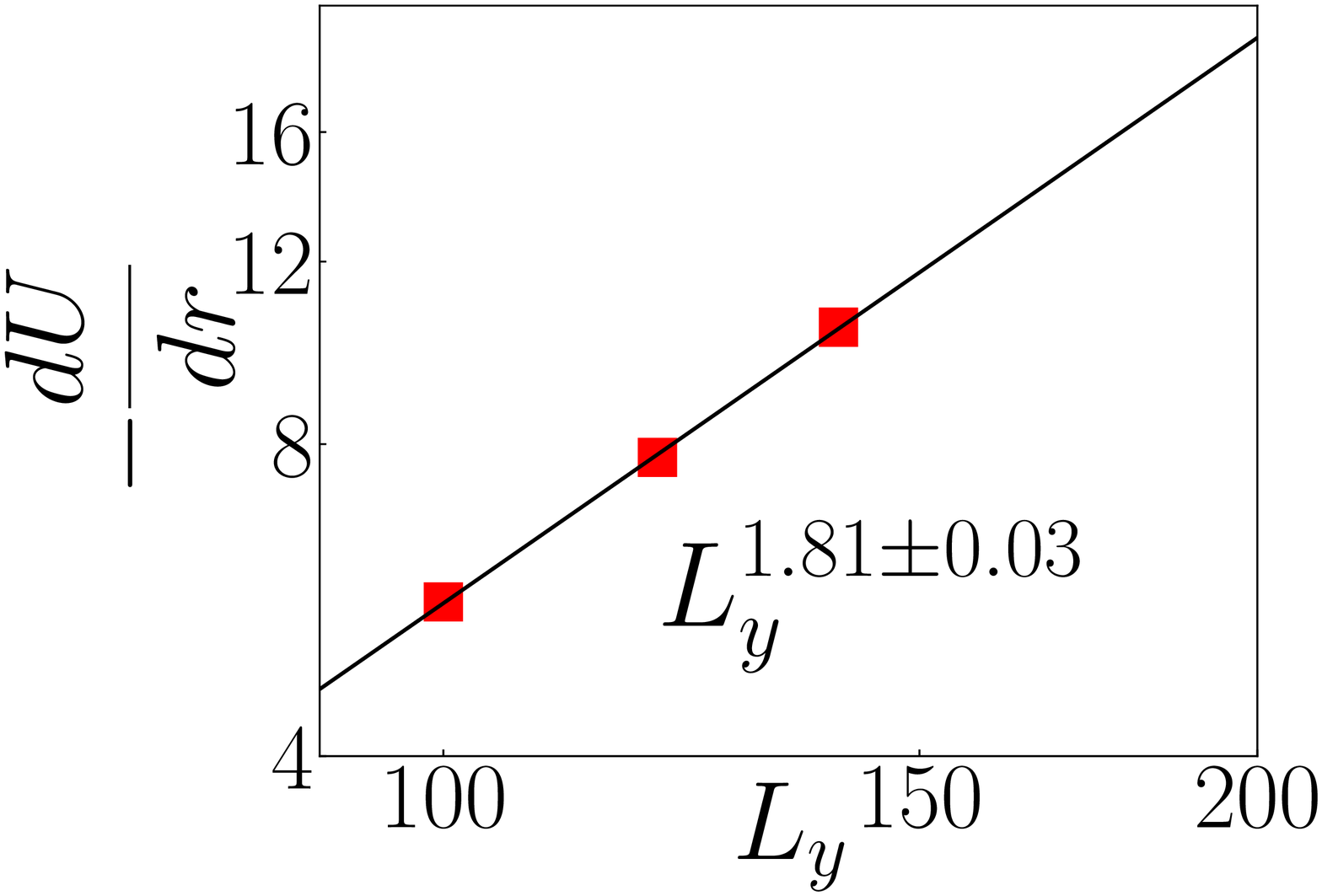}
}\\
\subfigure[$r=-2.6750$ ($r_c=-2.6743$) $\dot{\gamma}=0.01$]{
\includegraphics[width=6cm]{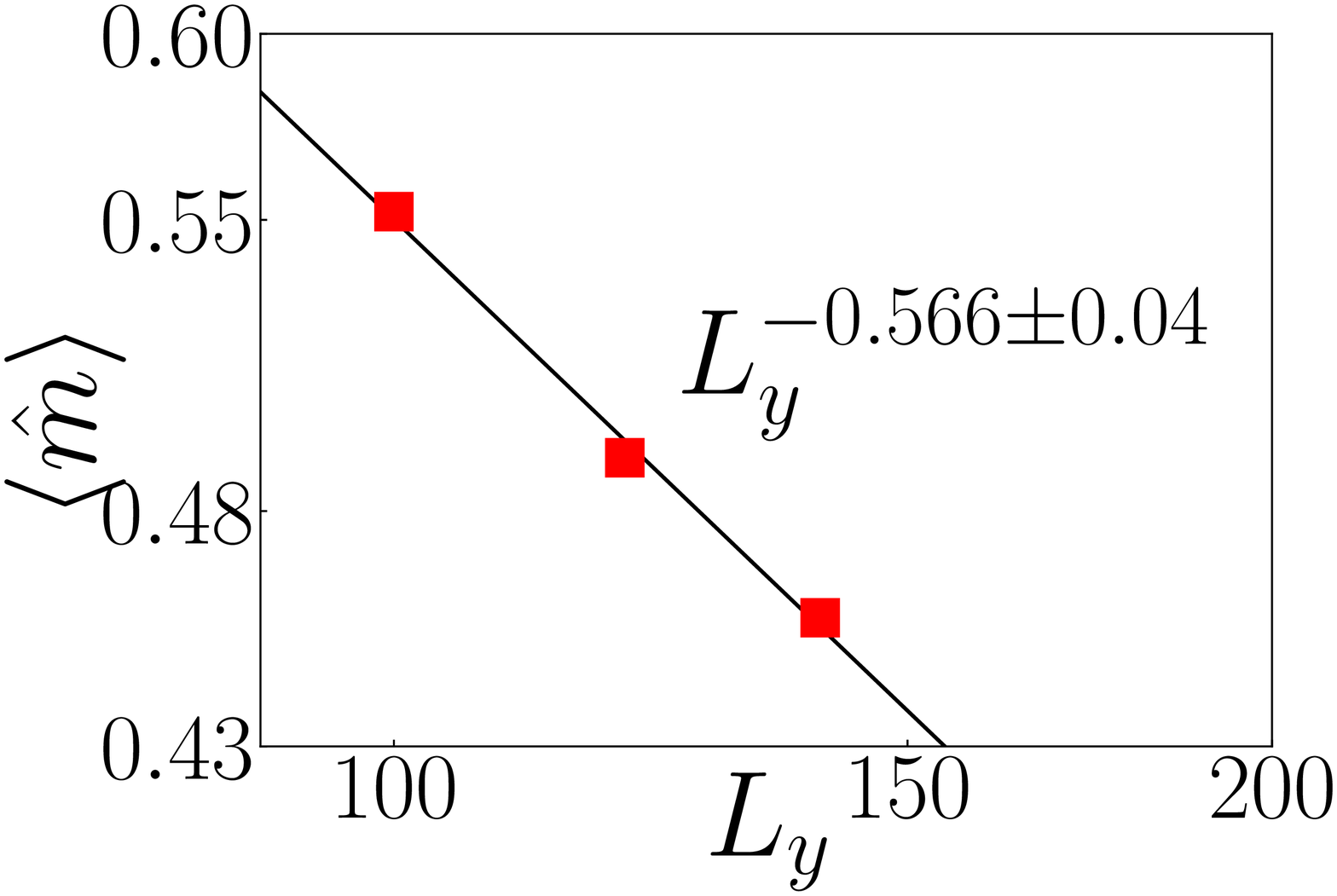} 
\includegraphics[width=6cm]{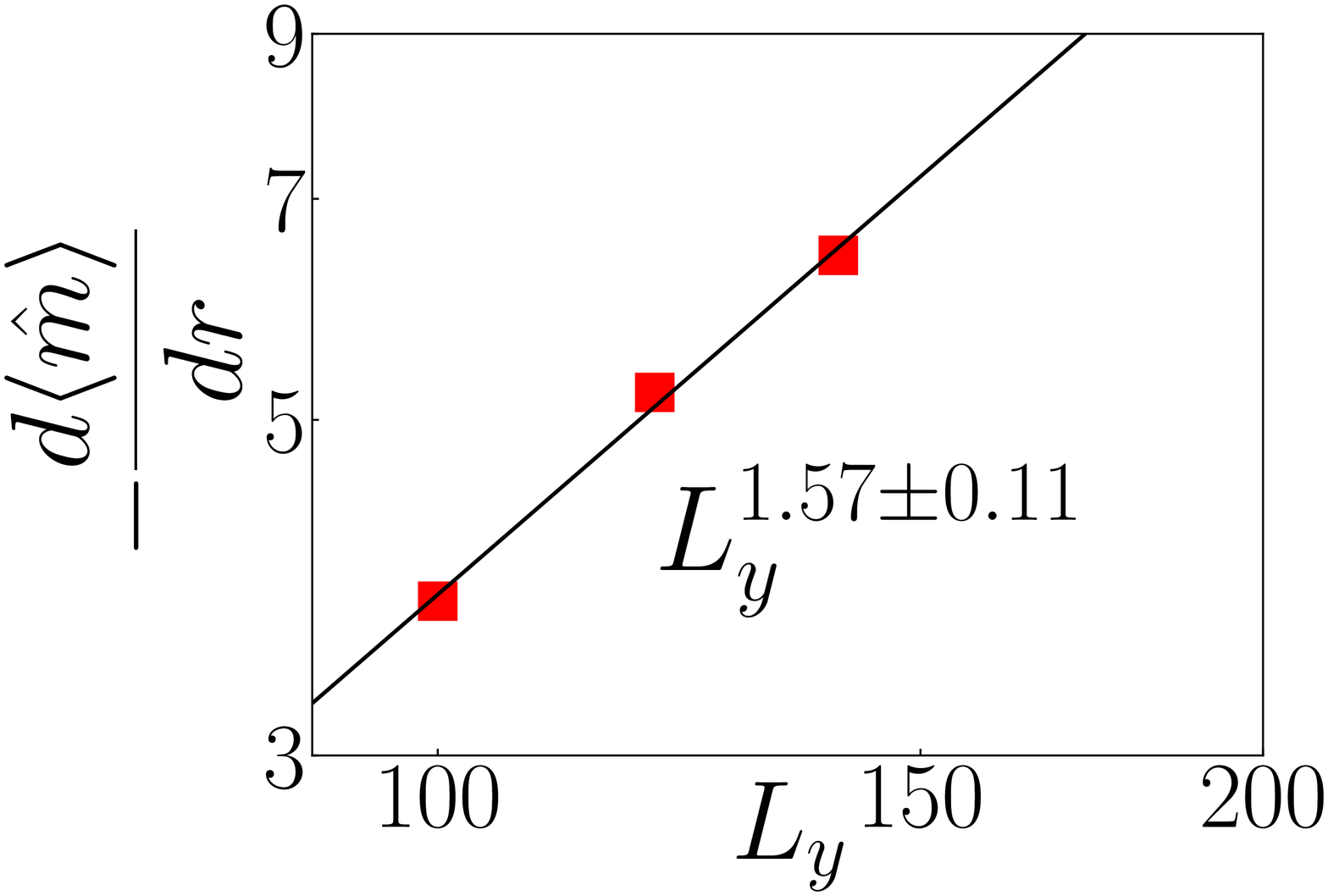}
\includegraphics[width=6cm]{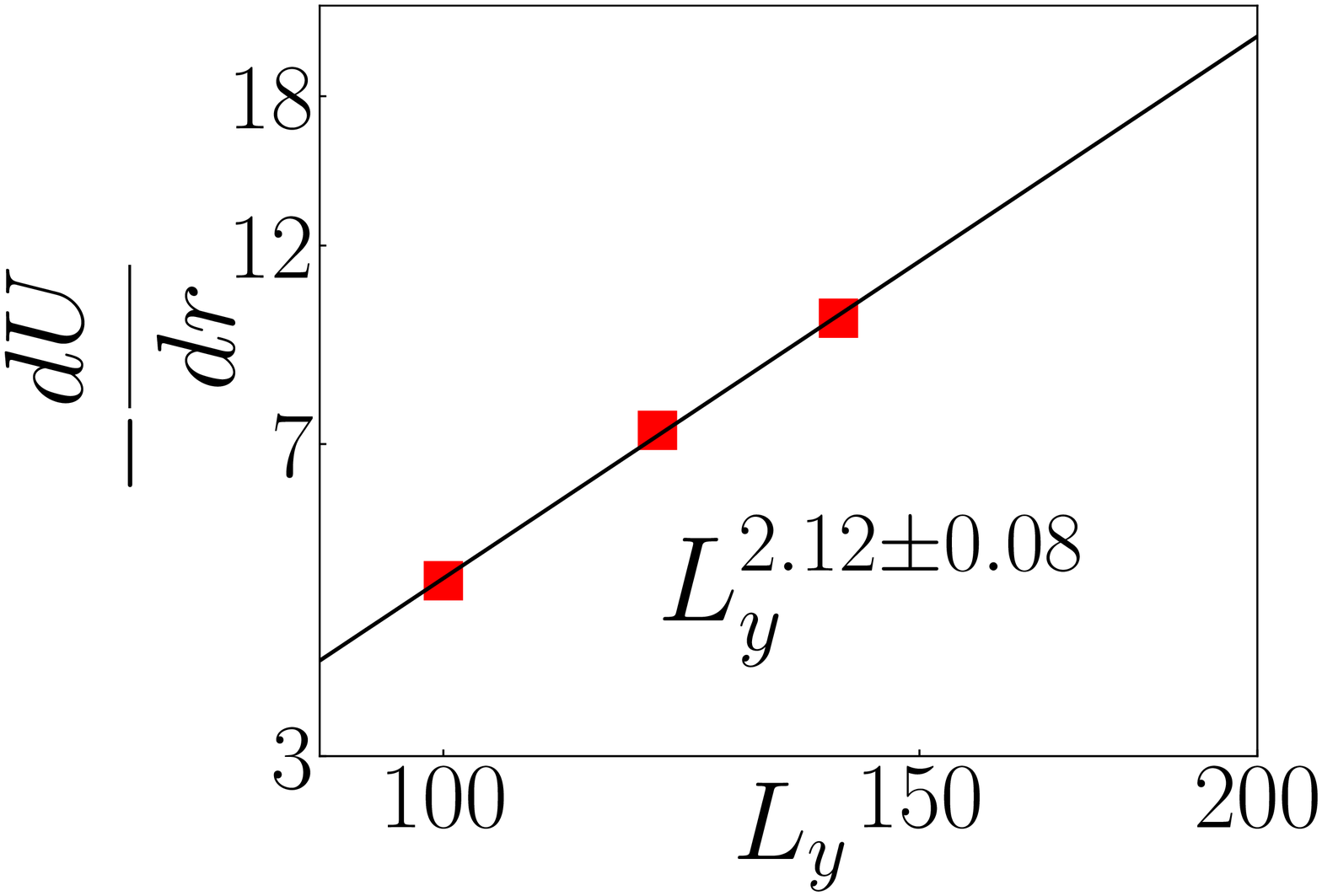} 
}
\vspace{-0.3cm}
\caption{(Color online) a log-log plot of the simulation data at the critical point. Left: $\langle \hat{m} \rangle$ versus $L_y$. Center: $d\langle\hat{m} \rangle_{h=0}/dr$ versus $L_y$. Right: $dU/dr$ versus $L_y$.}
\vspace{-0.3cm}
\label{fig:plot of scaling of magnetization}
\end{figure}

\subsection{Structure factor slightly above critical point}
We numerically calculate the structure factor slightly above the critical point and test whether the value of $z_x$ is equal to $3$. 
In Fig.~\ref{fig:plot of structure factor near criticality: sr5}, we show the structure factor $S(\bm{k})$ for $\dot{\gamma}=5.0$ and $r=-1.9255$ ($r_c=-1.9257\pm0.0002$). By using the fact that the structure factor $S(\bm{k})$ behaves as
\begin{eqnarray}
S(\bm{k}) = \frac{1}{C_\tau \tau^{\omega_\tau} + C_x k_x^{\omega_x} + C_y k_y^{\omega_y}}
\label{eq:structure factor: functional form}
\end{eqnarray}
slightly above the critical point, we fit the simulation data in $0.0<k_x \ {\rm or}\ k_y < 0.33$ and obtain $\omega_x=0.663 \pm 0.005$ and $\omega_y=1.98\pm0.01$. Since $\nu_x$ and $\nu_y$ are written as $\nu_i = \omega_{\tau}/\omega_i$, $z_x$ is calculated as $z_x=\nu_x/\nu_y=\omega_y/\omega_x=2.99\pm0.03$. The value $z_x=2.99$ is extremely close to $z_x=3$. Thus, we again confirm the validity of $z_x=3$. We also note that $\omega_x=0.663$ and $\omega_y=1.98$ are extremely close to that of linearized model, $\omega_x=2/3$ and $\omega_y=2$ (See Sec.~\ref{sec:fluctuations near criticality}).
\begin{figure}[t]
\centering
\includegraphics[width=8cm]{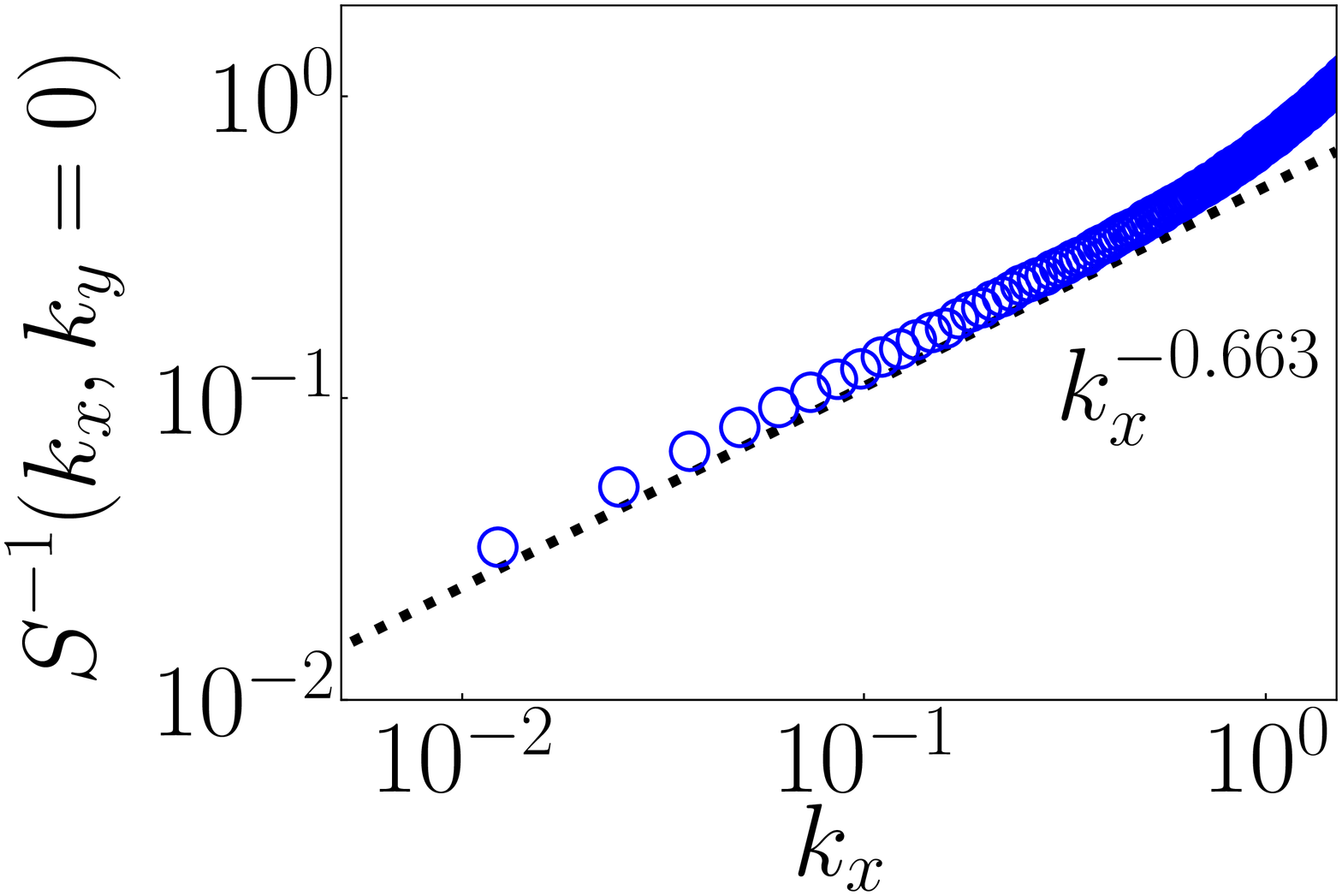} 
\includegraphics[width=8cm]{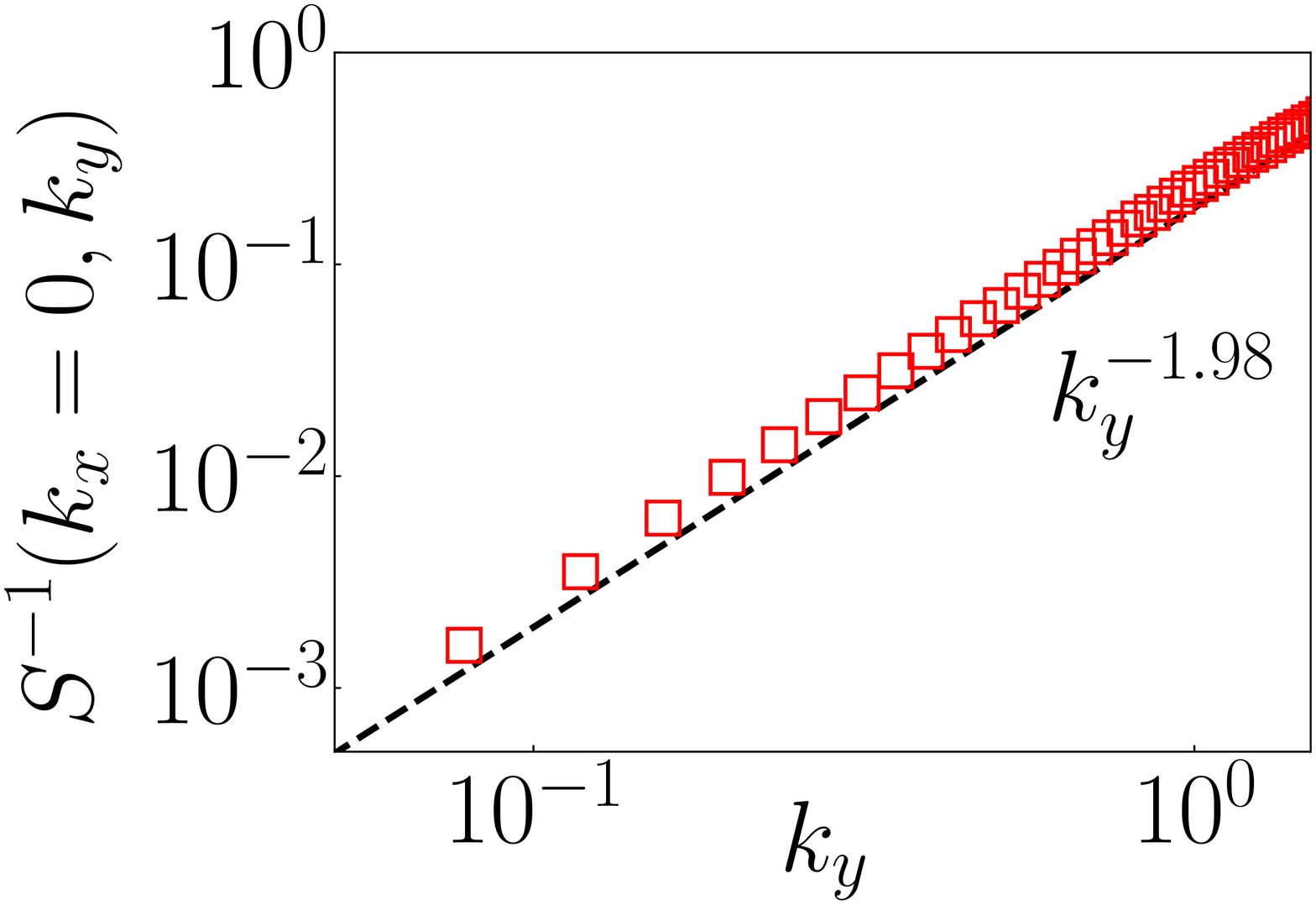}
\vspace{-0.3cm}
\caption{(Color online) Structure factor for $\dot{\gamma}=5.0$. $r$ is chosen as $r=-1.9255$, which is slightly above the critical point ($r_c=-1.9257\pm0.0002$). Left: $S^{-1}(k_x,k_y=0)$ versus $k_x$. Right: $S^{-1}(k_x=0,k_y)$ versus $k_y$.}
\vspace{-0.3cm}
\label{fig:plot of structure factor near criticality: sr5}
\end{figure}

The similar calculation is performed for $\dot{\gamma}=0.5$ and $0.01$. Here, we note that the $k_x$-dependence of $S^{-1}(k_x,k_y=0)$ crosses over from $k_x^{-2/3}$ to $k_x^{-2}$ at the length scale $l=\sqrt{\kappa\Gamma/\dot{\gamma}}$. This property is the same as the phase fluctuations explained in the main text. Because the length scale $l$ becomes larger as $\dot{\gamma}$ smaller, it is difficult to estimate the correct value of $\omega_x$ from the fitting of $S^{-1}(k_x,k_y=0)$. Then, instead of estimating $\omega_x$ and $\omega_y$ from the fitting, we compare the simulation data and the straight line with the theoretical values $\omega_x=2/3$ and $\omega_y=2$. The result is shown in Fig.~\ref{fig:plot of structure factor near criticality: sr smaller} and the good agreement is found between them, revealing that the structure factor is kept in the linearized form even for small $\dot{\gamma}$.
\begin{figure}[tb]
\centering
\subfigure[$\dot{\gamma}=0.5$, $r=-2.3280$ ($r_c=-2.3318$)]{
\includegraphics[width=6cm]{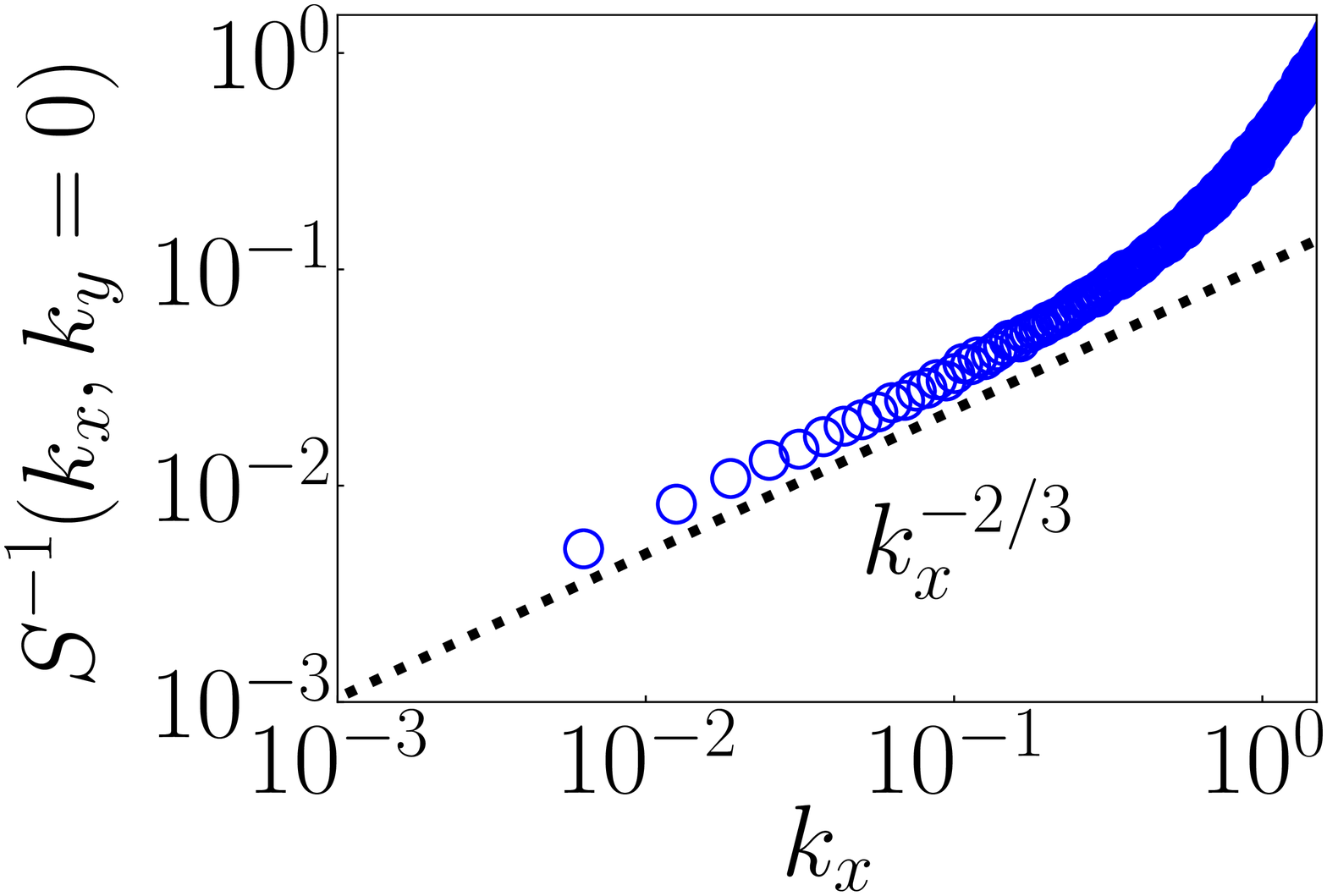} 
\includegraphics[width=6cm]{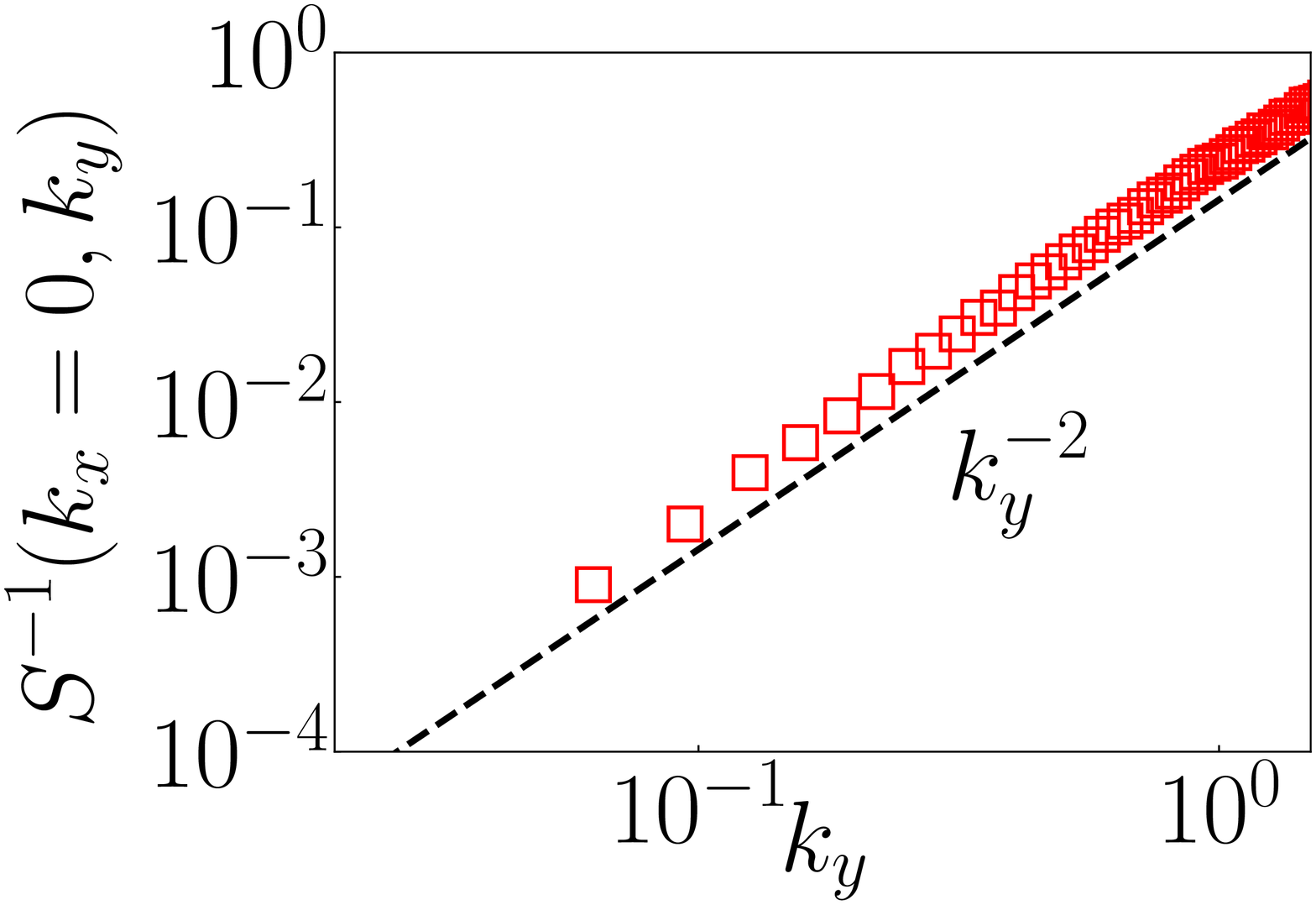}
}\\
\subfigure[$\dot{\gamma}=0.01$, $r=-2.6700$ ($r_c=-2.6743$)]{
\includegraphics[width=6cm]{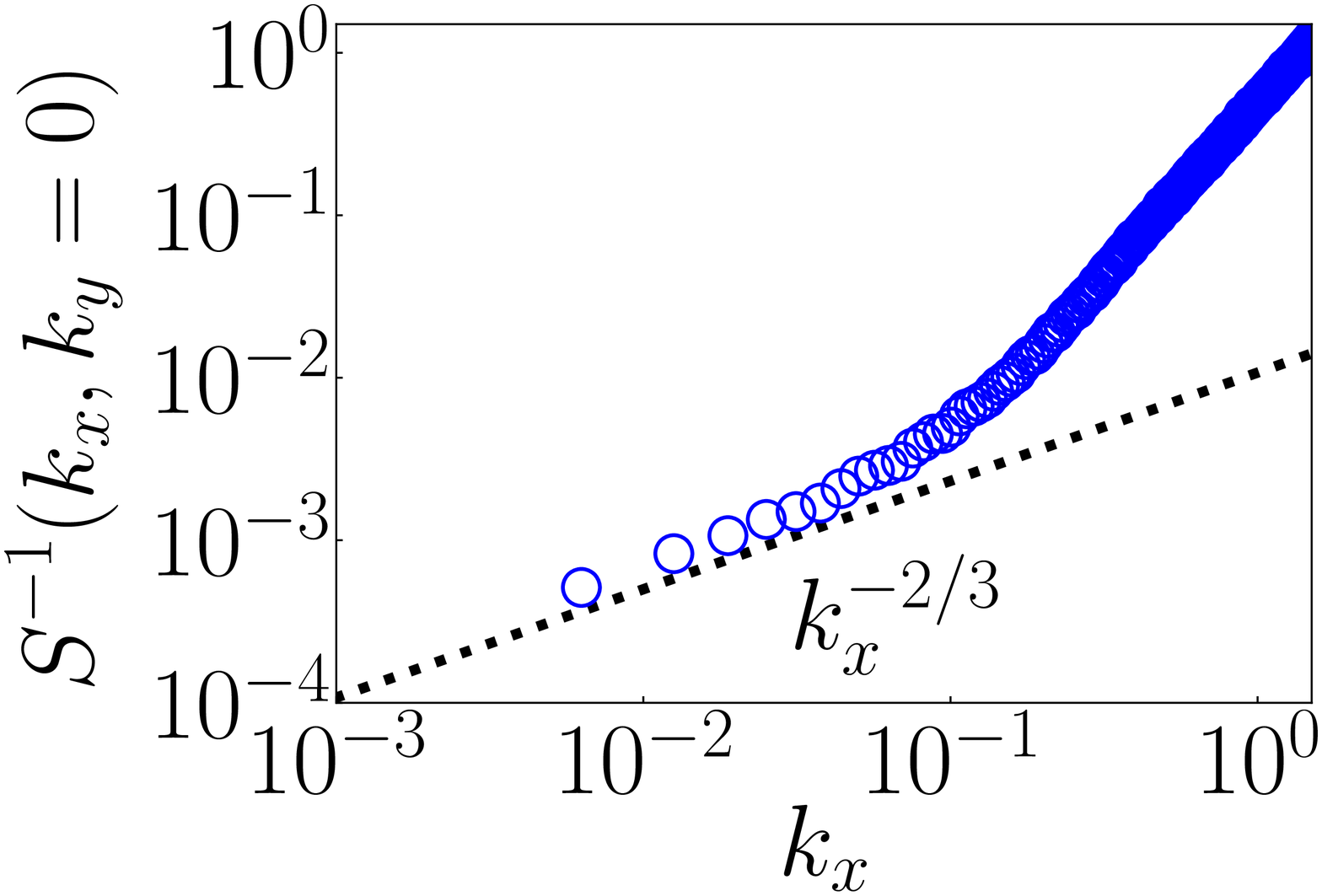}
\includegraphics[width=6cm]{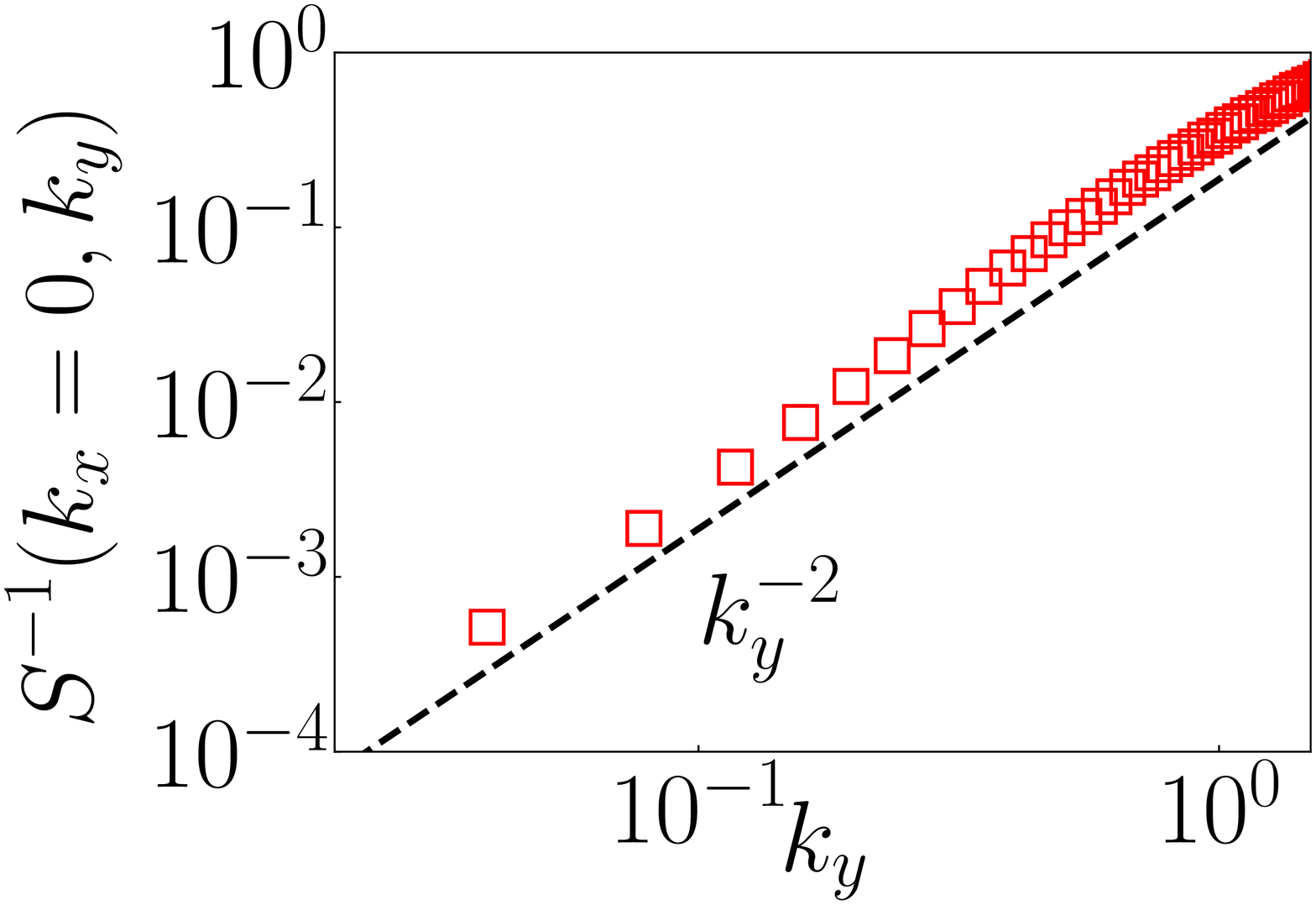} 
}
\vspace{-0.3cm}
\caption{(Color online) Same as Fig.~\ref{fig:plot of structure factor near criticality: sr5}, but with (a) $\dot{\gamma}=0.5$ and (b) $\dot{\gamma}=0.01$.}
\vspace{-0.3cm}
\label{fig:plot of structure factor near criticality: sr smaller}
\end{figure}

\subsection{Comparison with the previous studies}
To summarize our numerical result, (i) the value of the critical exponent $\beta$ is extremely close to the mean-field value for the large shear rate, (ii) it deviates from the mean-field theory when the shear rate becomes small, and (iii) $k_x^{-2/3}$ mode (i.e. $\omega_x=2/3$) is observed for all the shear rates we have examined.

Here, we compare our result with the ones previously obtained for the sheared Ising model. The pioneer theoretical analysis was carried out for the three-dimensional model H (in the classification of Hohenberg and Halperin) by Onuki and Kawasaki~\cite{smOnuki1979}. They applied the renormalization group method to the sheared system and showed that the critical exponent $\beta$ is given by the mean-field theory at sufficiently large shear rates. Apart from three-dimensional system, Hucht~\cite{smHucht2009} proposed the model that can be solved exactly in the limit of large shear rate and demonstrated that in this limit, $\beta$ is equal to the mean-field value even in the two-dimensional system. Some groups attempted to numerically verify this theoretical prediction~\cite{smChan1990,smWinter2010,smSaracco2009}. However, to our knowledge, there was no computational study that observes the mean-field behavior of $\beta$. For example, Chan and Lin reported $\beta = 0.38 \pm 0.05$~\cite{smChan1990}, Winter \textit{et al.} $\beta \approx 0.37$~\cite{smWinter2010}, and Saracco and Gonnella $\beta = 0.39 \pm 0.01$ (for the largest shear rate)~\cite{smSaracco2009}. Therefore, our study is the first to observe $\beta$ extremely close to the mean-field value for finite shear rate.

The behavior of $\beta$ for small shear rate is still a controversial problem. Our result indicates that it deviates from the mean-field value. However, we do not judge whether this deviation comes from the finite-size effects or remains in the large system-size limit.

\section{Kosterlitz--Thouless transition point in equilibrium}
We use the non-equilibrium relaxation method for determining the Kosterlitz--Thouless transition point $r_{KT}$ in equilibrium. It is the method that estimates the position of the critical point from the dynamical properties around the critical point. We below summarize the concrete procedure. See Ref.~\cite{smOzeki2003,smOzeki2007} for details of non-equilibrium relaxation method.

Let us consider the relaxation process of the magnetization $\langle\hat{m}_a\rangle (t)$ from the all-aligned state $\varphi_1(\bm{r})=1$ and $\varphi_2(\bm{r})=0$, where $\hat{m}_a$ is defined by
\begin{eqnarray}
\hat{m}_a = \frac{1}{L_xL_y} \int d^2\bm{r} \varphi_a(\bm{r}).
\end{eqnarray}
For the disordered state (i.e. $r>r_{KT}$), the magnetization $\langle\hat{m}_a\rangle (t)$ exhibits the exponential decay:
\begin{eqnarray}
\langle\hat{m}_x\rangle(t) \sim \exp\Big(-\frac{t}{\tau_{rel}(r)}\Big),
\end{eqnarray}
where $\tau_{rel}(r)$ is the relaxation time. The theoretical calculation predicts that the relaxation time $\tau_{rel}(r)$ diverges as $r \to r_{KT}+0$ in the form
\begin{eqnarray}
\tau_{rel}(r) = B \exp\Big(\frac{A}{\sqrt{r-r_{KT}}}\Big).
\label{eq:theoretical prediction of tau(r)}
\end{eqnarray}
Based on this property, we can estimate $r_{KT}$ from observing the divergence of relaxation time. 

In order to calculate the relaxation time $\tau_{rel}(r)$ in the numerical simulation, we use the dynamical scaling relation that holds near the critical point:
\begin{eqnarray}
\langle\hat{m}_x\rangle(t) = \tau_{rel}(r)^{\lambda} \tilde{m}\Big(\frac{t}{\tau_{rel}(r)}\Big),
\end{eqnarray}
where $\lambda$ is the dynamical exponent. Because $\lambda$ is a universal constant, we assume that $\lambda$ is given by the value, 0.068, obtained in the previous study~\cite{smOzeki2003}. We present the numerical calculation result of $\langle\hat{m}_x\rangle (t)$ and its scaling plot in Fig.~\ref{fig:relaxation of megnetization}. The system size is chosen as $L_x=L_y=512$, and we take an ensemble average over $120$ noise realizations. While $\tau_{rel}(r)$ hardly depends on the system size far from the transition point $r_{KT}$, a larger system size is required to exactly measure $\tau_{rel}(r)$ very near the transition point $r_{KT}$. To depict the scaling plot, we choose the magnetization curve with $r=-2.75$ as the reference curve, specifically, $\tau_{rel}(r=-2.75)$ is fixed at $1.0$, and fit the magnetization curves with $r=-2.895,-2.89,-2.885,-2.88,-2.875,-2.87,-2.86,-2.85,-2.84,-2.83,-2.82,-2.81$ to the reference curve. The best-fit parameter $\tau_{rel}$ is depicted in Fig.~\ref{fig:relaxation time} as a function of $r$. It is well fitted by Eq.~(\ref{eq:theoretical prediction of tau(r)}) with $A=5.6190\pm0.4300$, $\log B=-10.820\pm0.6868$ and $r_{KT} =  -3.0204\pm0.0087$. Accordingly, the Kosterlitz--Thouless transition point is estimated as $r_{KT} =  -3.0204\pm0.0087$.

\begin{figure}[t]
\centering
\includegraphics[width=8cm]{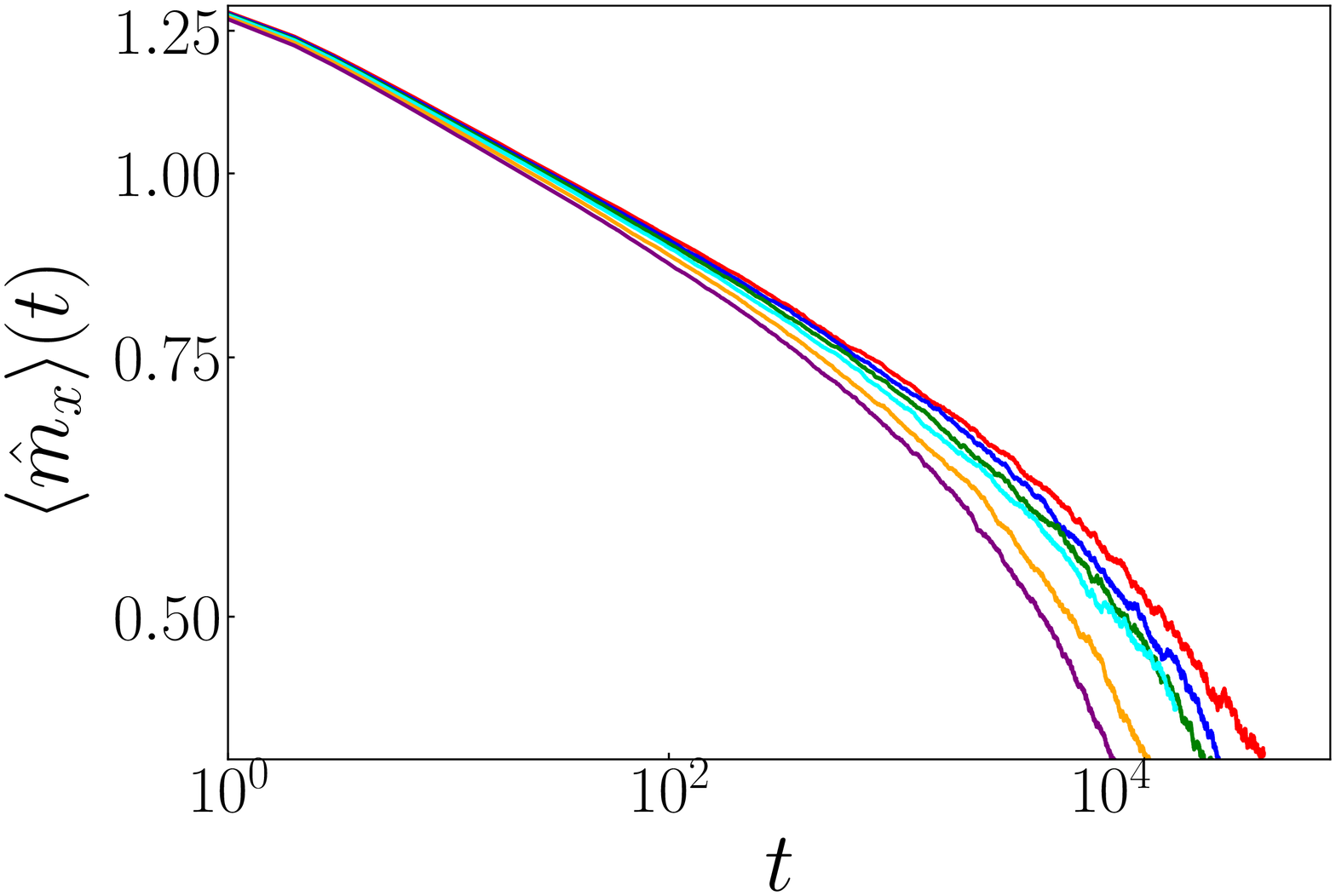} 
\includegraphics[width=8cm]{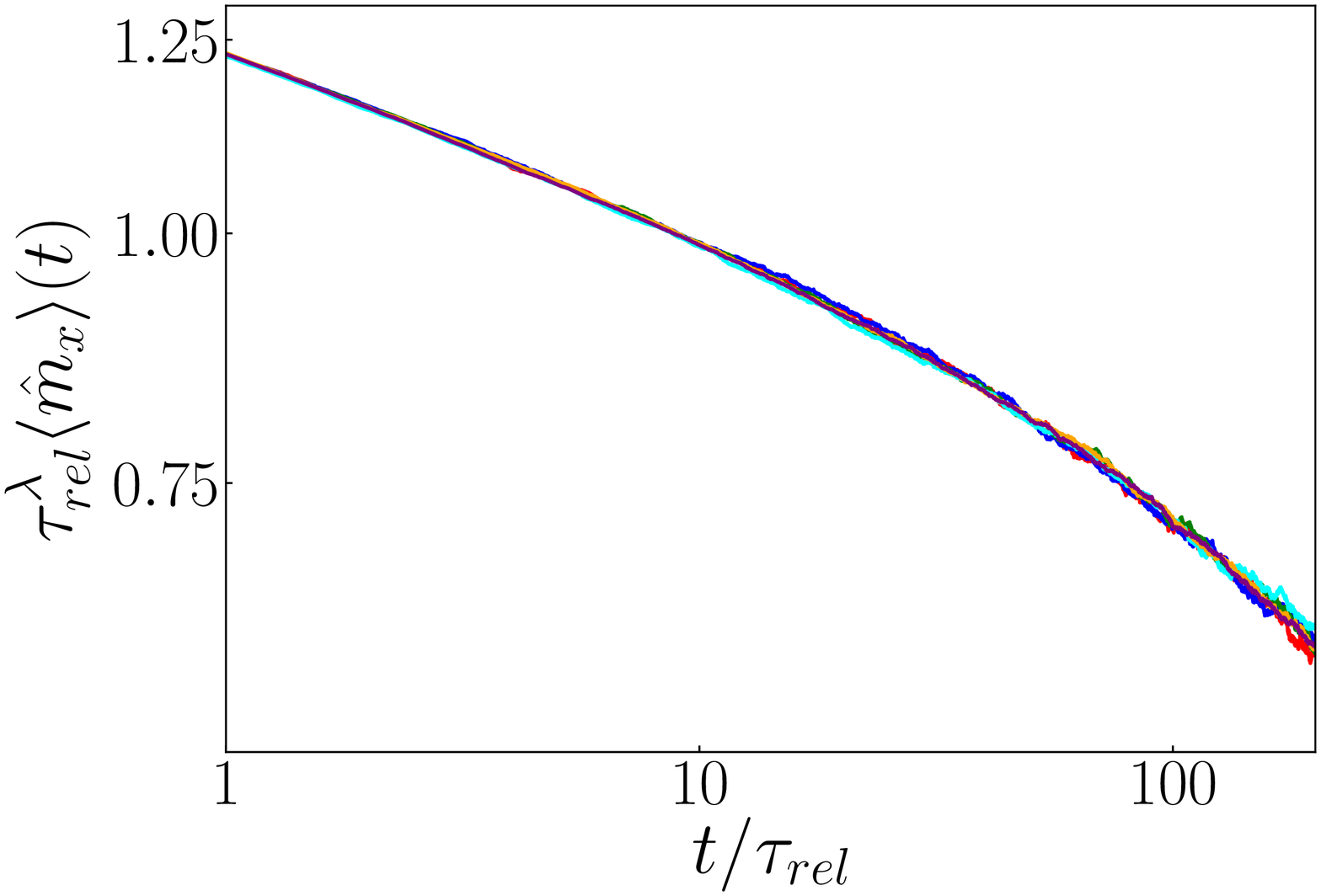} \\
\caption{(Color online) Left: relaxation of magnetization from the all-aligned state for $r=-2.895,-2.89,-2.885,-2.88,-2.87,-2.86$ with a log-log plot. Right: $\tau_{rel}^{\lambda}\langle\hat{m}_x\rangle (t)$ versus $t/\tau_{rel}$ calculated from the left figure. Each curve is shifted by $\lambda=0.068$ and $\tau_{rel} = 158.0,110.0,84.0,69.0,37.0,25.0$.}
\label{fig:relaxation of megnetization}
\end{figure}
\begin{figure}[t]
\centering
\includegraphics[width=8cm]{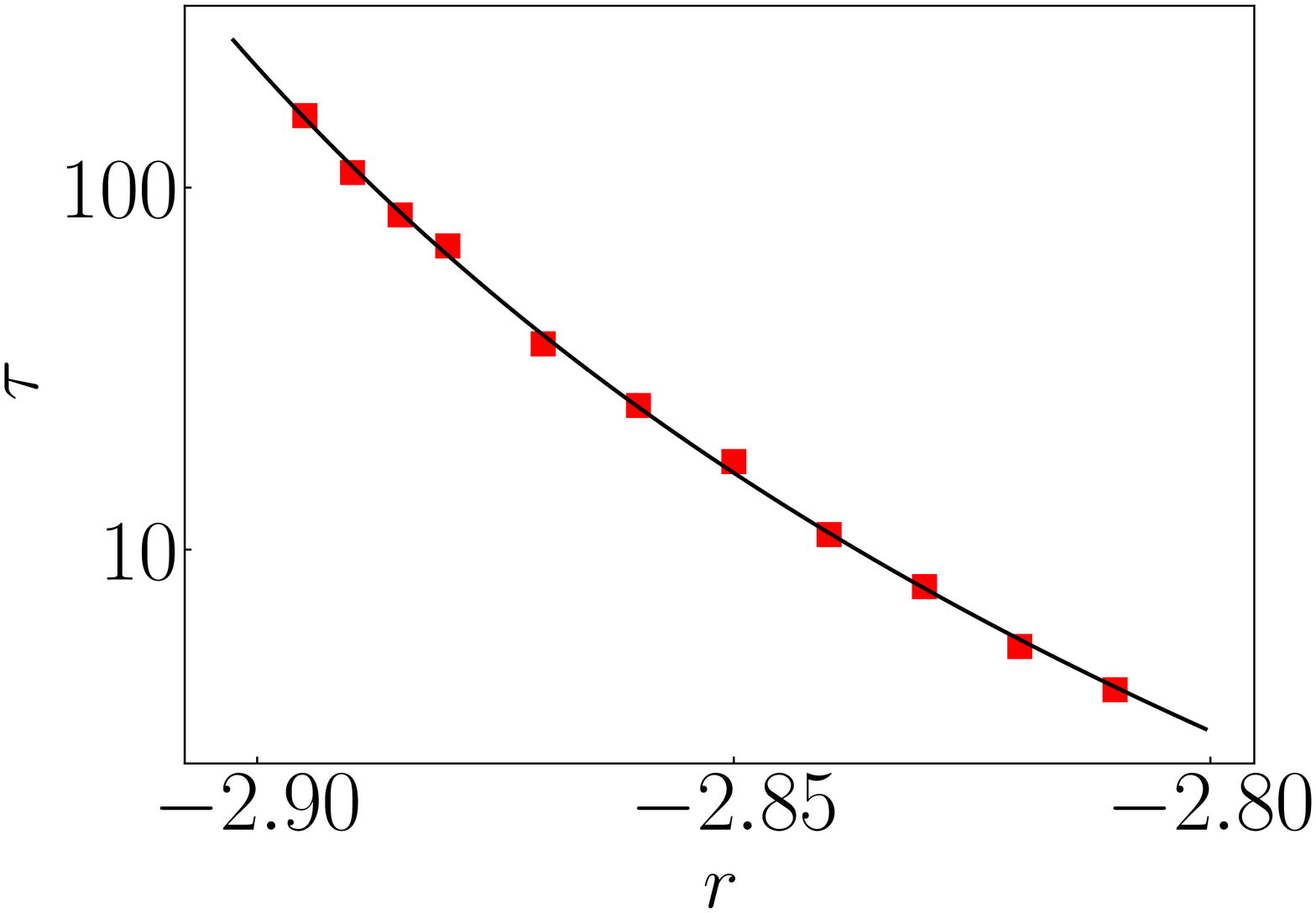} 
\includegraphics[width=8cm]{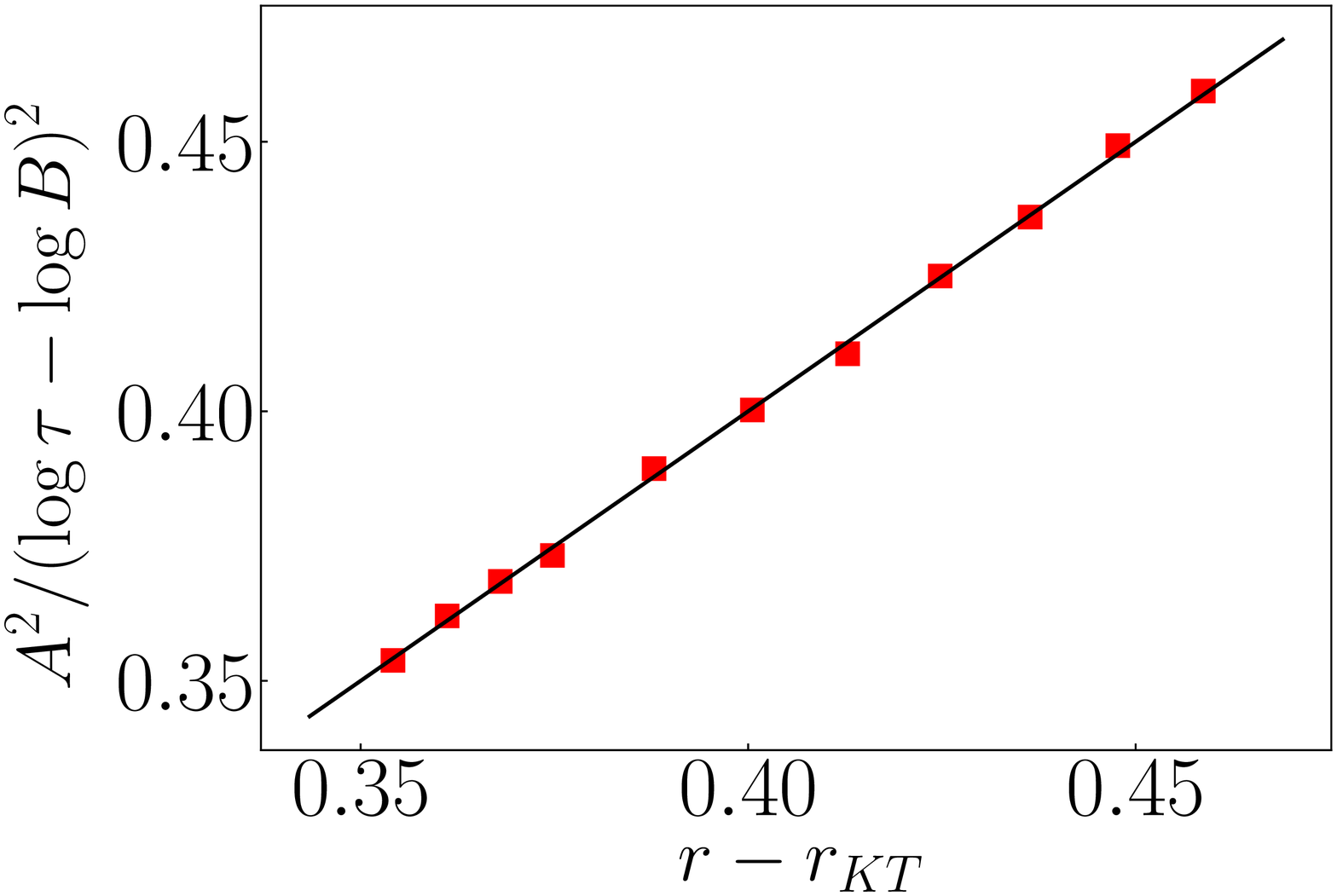} 
\caption{(Color online) Relaxation time $\tau_{rel}$ as a function of $r$. Left: $\tau_{rel}$ versus $r$. Right: $A^2/(\log\tau_{rel}-\log B)^2$ versus $r-r_{KT}$. The black solid curve represents Eq.~(\ref{eq:theoretical prediction of tau(r)}) with $A=5.6190$, $\log B=-10.820$ and $r_{KT} = -3.0204$. }
\label{fig:relaxation time}
\end{figure}

Furthermore, we measure the helicity modulus $\Upsilon$ in equilibrium state~\cite{smWeber1988}, which is defined as follows~\cite{smFisher1973,smOhta1979}. Let us consider the twisted periodic boundary condition along $x$ direction
\begin{eqnarray}
\bm{\varphi}(x+L,y,t) = R(\Delta) \bm{\varphi}(x,y,t),
\label{eq: twisted periodic boundary condition}
\end{eqnarray}
where $R(\Delta)$ is the rotation matrix
\begin{eqnarray}
R(\Delta) = \begin{pmatrix}
\cos \Delta & -\sin \Delta \\
\sin \Delta & \cos \Delta
\end{pmatrix}.
\label{eq:rotation matrix}
\end{eqnarray}
The free energy depends on the twisted angle $\Delta$. We write it as $F(T,\Delta,L)$, where $T$ is the temperature and $L$ is the system size $L_x=L_y$. The free energy $F(T,\Delta,L)$ is expanded in the form
\begin{eqnarray}
F(T,\Delta,L) = F(T,0,L) + \frac{\partial F(T,\Delta,L) }{\partial \Delta}\Big|_{\Delta=0} \Delta + \frac{1}{2}\frac{\partial^2 F(T,\Delta,L) }{\partial \Delta^2}\Big|_{\Delta=0} \Delta^2 + \cdots ,
\end{eqnarray}
where $F(T,0,L)$ is the free energy under the standard periodic boundary condition. Because the system is invariant under $\Delta \to -\Delta$, we have
\begin{eqnarray}
\frac{\partial F(T,\Delta,L) }{\partial \Delta}\Big|_{\Delta=0} = 0.
\end{eqnarray}
Furthermore, $F(T,\Delta,L) \geq F(T,0,L)$ holds because the global minimum of free energy corresponds to the non-twisted state. We thus obtain
\begin{eqnarray}
\frac{\partial^2 F(T,\Delta,L) }{\partial \Delta^2}\Big|_{\Delta=0} \geq 0.
\end{eqnarray}
Based on these properties, the helicity modulus $\Upsilon$ is defined as
\begin{eqnarray}
\Upsilon(L) = \frac{\partial^2 F(T,\Delta,L) }{\partial \Delta^2}\Big|_{\Delta=0}.
\label{eq:definition of helicity modulus}
\end{eqnarray}

We here introduce $\Upsilon_{\infty} \equiv \lim_{L \to \infty}\Upsilon(L)$. Intensive studies revealed that $\Upsilon_{\infty}$ jumps at the Kosterlitz--Thouless transition point $r_{KT}$ from zero (in disordered state) to $2T/\pi$ (in quasi-long-range ordered state)~\cite{smWeber1988,smSchultka1994,smOlsson1995}, and that the finite-size corrections are given by
\begin{eqnarray}
\Upsilon(L) = \frac{2T}{\pi}\Big(1 + \frac{1}{2} \frac{1}{\log L + {\rm const}}\Big),
\label{eq:first order correction to helicity modulus}
\end{eqnarray}
which was derived by Weber and Minnhagen~\cite{smWeber1988}. Note that higher-order corrections were discussed in Ref.~\cite{smHasenbusch2005}. 

We test Eq.~(\ref{eq:first order correction to helicity modulus}) near $r=-3.0204$ and confirm the validity of the transition point obtained by the non-equilibrium relaxation method. Fig.~\ref{fig:helicity modulus} presents the helicity modulus calculated in the numerical simulations. The system size is chosen as $L=32,64,128,256$. We take an average over $10^6$ different times at $t=100i\delta t$ for $8$ noise realizations. We will later explain the microscopic expression used to calculate the helicity modulus.
\begin{figure}[t]
\centering
\includegraphics[width=8.6cm]{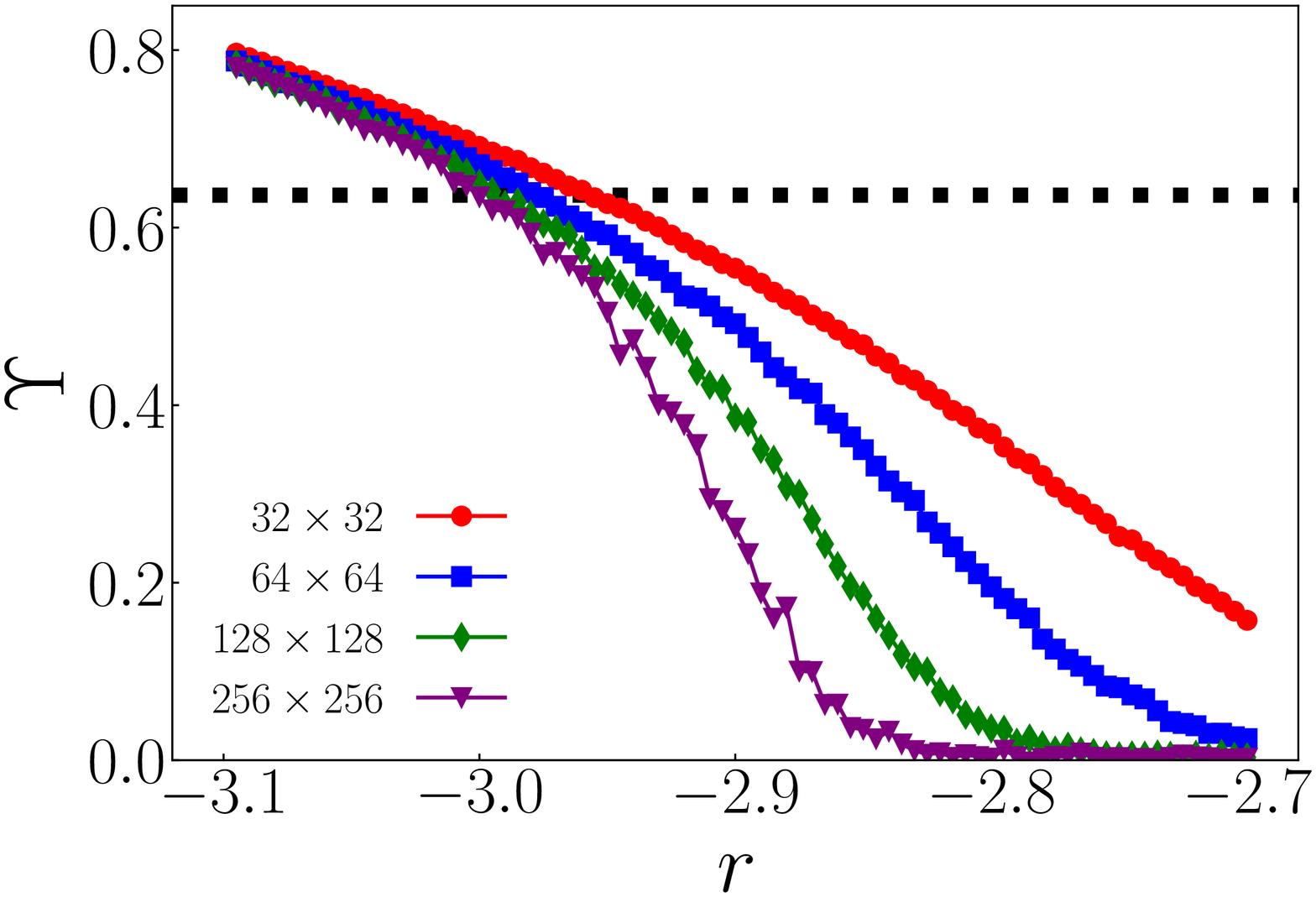} 
\includegraphics[width=8.6cm]{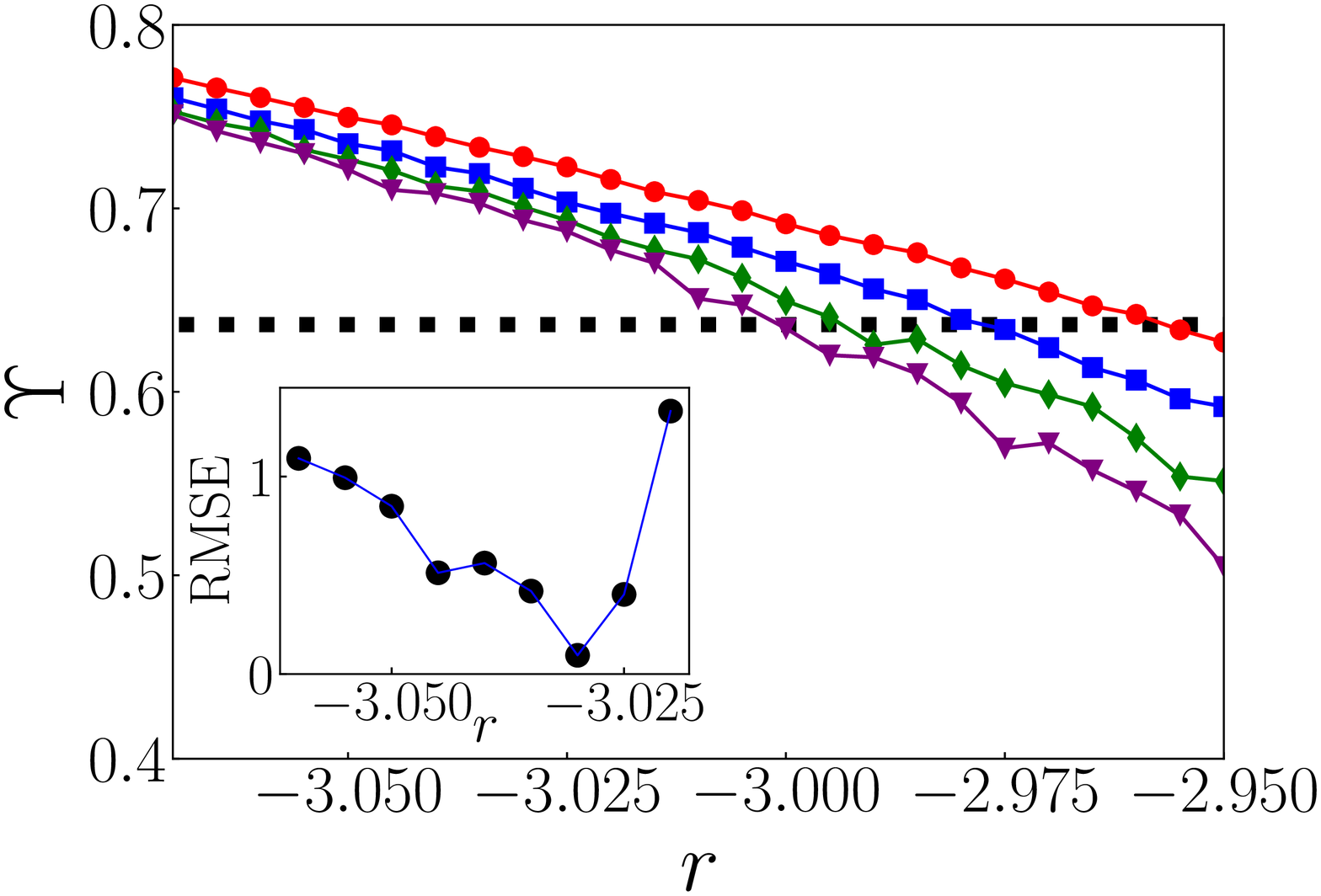} 
\caption{(Color online) Plot of helicity modulus for four different system sizes. Left: $\Upsilon$ versus $r$. Right: zoom around $r=-3.00$. Inset: root mean square error (RMSE) of fit to Eq.~(\ref{eq:the finite size relation of the helicity modulus: another}) at each $r$. The minimum point gives the Kosterlitz--Thouless transition point. }
\label{fig:helicity modulus}
\end{figure}
In the left side of Fig.~\ref{fig:helicity modulus}, we observe the onset of the helicity modulus from zero to the finite value. As shown in Eq.~(\ref{eq:first order correction to helicity modulus}), the helicity modulus at $r=r_{KT}$ approaches $2T/\pi$ in the limit $L\to \infty$, which is depicted by the black dotted line. Because the helicity modulus for $L=256$ takes a value close to $2T/\pi$ at $r=3.0$, we plot the zoom of this region in the right side of Fig.~\ref{fig:helicity modulus}.

Then, we fit the simulation data at each $r$ to Eq.~(\ref{eq:first order correction to helicity modulus}) by using the least squares method. For this purpose, we rewrite Eq.~(\ref{eq:first order correction to helicity modulus}) into 
\begin{eqnarray}
\Big(\Upsilon(L) \frac{\pi}{2T} - 1\Big)^{-1} =  2 \big(\log L + {\rm const}\big),
\label{eq:the finite size relation of the helicity modulus: another}
\end{eqnarray}
and const is treated as a free parameter. The root mean square error (RMSE) of fit to Eq.~(\ref{eq:the finite size relation of the helicity modulus: another}) is presented in the inset of the right side of Fig.~\ref{fig:helicity modulus}. It takes a minimum at $r=-3.03$, which means that the Kosterlitz--Thouless transition point $r_{KT}$ is located near $r=-3.03$. We show the simulation data (red square) and the best-fit curve (black solid) at $r=-3.03$ in Fig.~\ref{fig:helicity modulus: best fit}. To make it easier to see, it is organized in the form of Eq.~(\ref{eq:the finite size relation of the helicity modulus: another}). From this figure, we confirm the validity of our fitting result.
\begin{figure}[ht]
\centering
\includegraphics[width=8.6cm]{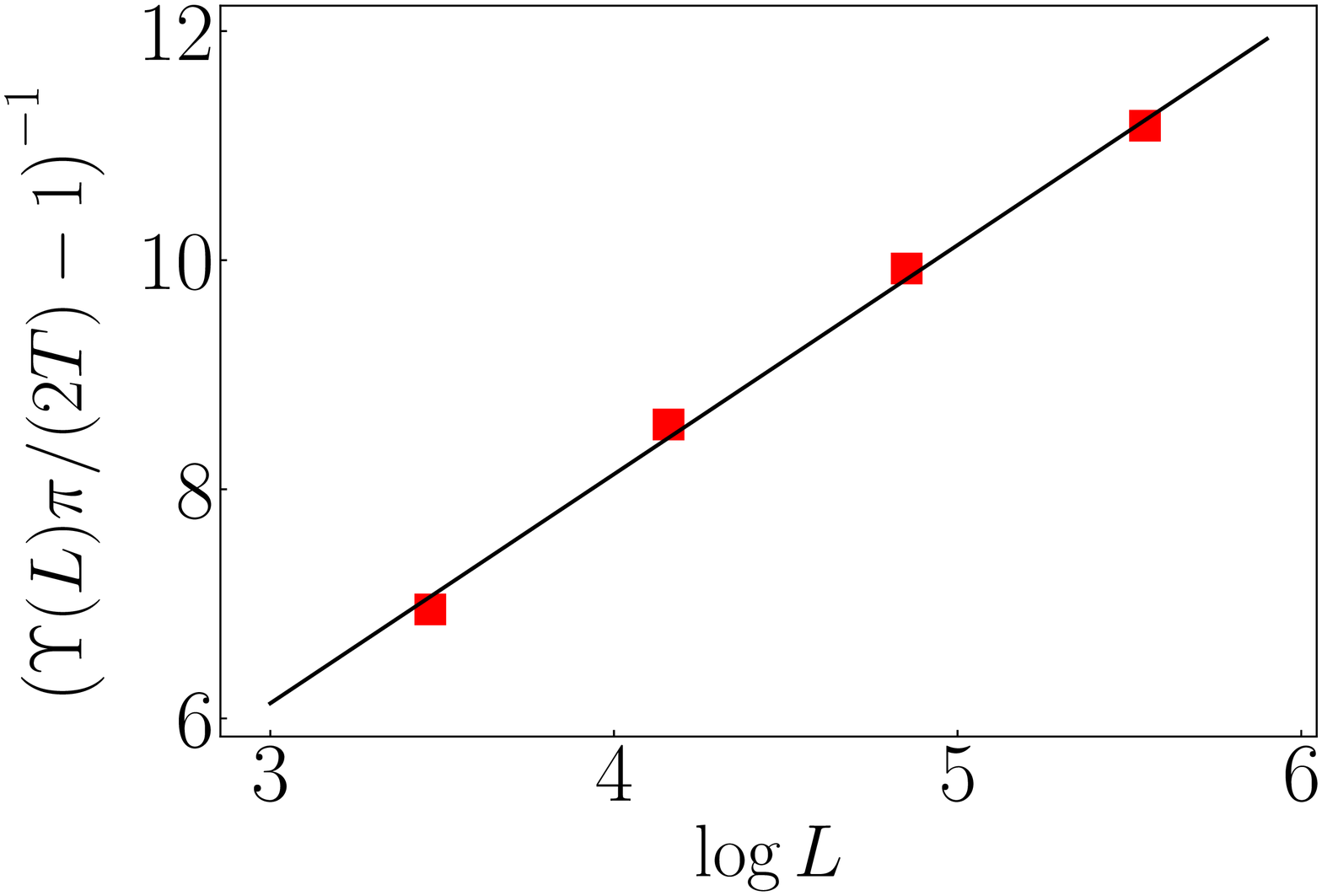} 
\caption{(Color online) $\Upsilon$ versus $L$ at $r=-3.03$.}
\label{fig:helicity modulus: best fit}
\end{figure}

The calculation result using the helicity modulus is in reasonable agreement with that by the non-equilibrium measurement method. This consistency justifies the non-equilibrium measurement method and we conclude that $r_{KT} = -3.0204\pm0.0087$.

\subsubsection*{Microscopic expression of  helicity modulus}
To calculate the helicity modulus in numerical simulations, we derive a microscopic expression of helicity modulus. We start with the spatially-discretized Landau free energy:
\begin{eqnarray}
\hspace{-0.3cm}\Phi[\bm{\varphi}] = (\delta x)(\delta y) \sum_{i_x,i_y} \Big\{\frac{\kappa}{2}\sum_{a=1}^2\Big(\frac{\varphi_a^{i_x+1,i_y}-\varphi_a^{i_x,i_y}}{\delta x}\Big)^2 + \frac{\kappa}{2}\sum_{a=1}^2\Big(\frac{\varphi_a^{i_x,i_y+1}-\varphi_a^{i_x,i_y}}{\delta y}\Big)^2 + \frac{r}{2}|\bm{\varphi}^{i_x,i_y}|^2 + \frac{u}{4}\big(|\bm{\varphi}^{i_x,i_y}|^2\big)^2 \Big\},
\end{eqnarray} 
where $\delta x$ and $\delta y$ are the space interval. This Landau free energy yields Eq.~(3) in the continuum limit.

Instead of considering the system under the twisted periodic boundary condition Eq.~(\ref{eq: twisted periodic boundary condition}), we introduce the twisted Landau free energy
\begin{eqnarray}
\Phi[\bm{\varphi};\Delta] = (\delta x)(\delta y) \sum_{i_x,i_y} & & \Big\{\frac{\kappa}{2}\sum_{a=1}^2\Big(\frac{\psi_a^{i_x+1,i_y}(\Delta)-\psi_a^{i_x,i_y}(\Delta)}{\delta x}\Big)^2 + \frac{\kappa}{2}\sum_{a=1}^2\Big(\frac{\psi_a^{i_x,i_y+1}(\Delta)-\psi_a^{i_x,i_y}(\Delta)}{\delta y}\Big)^2 \nonumber \\
&+& \frac{r}{2}|\bm{\psi}^{i_x,i_y}(\Delta)|^2 + \frac{u}{4}\big(|\bm{\psi}^{i_x,i_y}(\Delta)|^2\big)^2 \Big\}
\end{eqnarray}
with
\begin{eqnarray}
\bm{\psi}^{i_x,i_y}(\Delta) = R\Big(\frac{\Delta}{L}i_x \delta x\Big)\bm{\varphi}^{i_x,i_y},
\end{eqnarray}
and study this system under the standard periodic boundary condition. These two systems are equivalent to each other with respect to thermodynamic properties.

The free energy is given by
\begin{eqnarray}
F(T,\Delta,L) = - \frac{1}{T} \log \int \Big(\prod_{i_x,i_y} d^2 \bm{\varphi}^{i_x,i_y}\Big) e^{-\Phi[\bm{\varphi};\Delta]/T}.
\label{eq:the free energy of the twisted system; another}
\end{eqnarray}
By substituting Eq.~(\ref{eq:the free energy of the twisted system; another}) into the definition of the helicity modulus $\Upsilon$, Eq.~(\ref{eq:definition of helicity modulus}), we obtain the microscopic expression of the helicity modulus:
\begin{eqnarray}
\Upsilon &=& - \frac{\kappa}{T} \Big(\frac{\delta x}{L}\Big)^2(\delta y)^2 \Big\{\Big\langle \Big(\sum_{i_x,i_y} \Big[\varphi_1^{i_x,i_y} \varphi_2^{i_x+1,i_y} - \varphi_2^{i_x,i_y}\varphi_1^{i_x+1,i_y}\Big] \Big)^2\Big\rangle  -\Big\langle \sum_{i_x,i_y} \Big[\varphi_1^{i_x,i_y} \varphi_2^{i_x+1,i_y} - \varphi_2^{i_x,i_y}\varphi_1^{i_x+1,i_y}\Big] \Big\rangle^2 \Big\}\nonumber \\
&+&  \Big(\frac{\delta x}{L}\Big)^2(\delta y)  \Big\langle\sum_{i_x,i_y} \Big[\varphi_1^{i_x,i_y} \varphi_1^{i_x+1,i_y} + \varphi_2^{i_x,i_y} \varphi_2^{i_x+1,i_y} \Big] \Big\rangle.
\end{eqnarray}
The numerical results in the previous subsection were obtained by using this expression.

%

\end{widetext}

\end{document}